\documentclass[10pt,journal]{IEEEtran}
\usepackage{cite}
\usepackage{amsmath,amssymb,amsfonts}
\usepackage{algorithmic}
\usepackage{graphicx}
\usepackage{textcomp}
\def\BibTeX{{\rm B\kern-.05em{\sc i\kern-.025em b}\kern-.08em
    T\kern-.1667em\lower.7ex\hbox{E}\kern-.125emX}}
\usepackage{mathtools}
\usepackage{bbm}
\usepackage{graphicx} 
\usepackage{thmtools}
\usepackage{flushend}
\usepackage{tikz}
\usepackage{cuted}
\usepackage{arydshln}
\usepackage{dblfloatfix}
\usepackage{xfrac}
\usepackage{float}

\usepackage{tikz}
\usepackage{tkz-graph}
\usepackage{graphicx}      
\usepackage{pgfplots} 
\pgfplotsset{compat=newest} 
\pgfplotsset{plot coordinates/math parser=false}
\newlength\figureheight
\newlength\figurewidth

\usepackage{pgfplots}
  \pgfplotsset{compat=newest}
  \pgfplotsset{plot coordinates/math parser=false}
  \usetikzlibrary{plotmarks}
  \usetikzlibrary{arrows.meta}
  \usepgfplotslibrary{patchplots}
  \usepackage{grffile}
  \usepackage{amsmath}

\newcommand{\rline}{{\mathbb R}}

\newcommand{\bbm}[1]{\left[\begin{matrix} #1 \end{matrix}\right]}
\newcommand{\sbm}[1]{\left[\begin{smallmatrix} #1
   \end{smallmatrix}\right]}

\newcommand{\rfb}[1]{\mbox{\rm
   (\ref{#1})}\ifx\undefined\stillediting\else:\fbox{$#1$}\fi}

\newenvironment{proof}{\vspace{.1cm}\noindent{\sc
    Proof.}\hspace{0.10cm}\,\,}{$\hfill\Box$\vspace{.3cm}} 

\newtheorem{theorem}{Theorem}[section] 
\newtheorem{definition}[theorem]{Definition}

\newtheorem{proposition}[theorem]{Proposition} 
\newtheorem{corollary}[theorem]{Corollary}
 
\newtheorem{example}[theorem]{Example}
\newtheorem{remark}[theorem]{Remark}
 
\renewcommand\thmcontinues[1]{Continued}

\makeatletter
\def\endthebibliography{%
	\def\@noitemerr{\@latex@warning{Empty `thebibliography' environment}}%
	\endlist
}

\begin{document}

\title{Kron-based Model-order Reduction of Open Mass-action Kinetics Chemical Reaction Networks}

\author{Mohamad Agung Prawira Negara, 
Azka M. Burohman,
Bayu Jayawardhana, 
\thanks{The authors are with the Engineering and Technology Institute Groningen, Faculty of Science and Engineering, University of Groningen, 9747AG Groningen, The Netherlands {\tt\small (m.a.prawira.negara@rug.nl, a.m.burohman@rug.nl, b.jayawardhana@rug.nl}).}
\thanks{M.A. Prawira Negara is also with Electrical Engineering, Faculty of Engineering, University of Jember, Jl. Kalimantan Tegalboto No.37, Krajan Timur, Sumbersari, Kec. Sumbersari, Kabupaten Jember, Jawa Timur 68121, Indonesia. {\tt\small magungpn@unej.ac.id}.}
\thanks{A.M. Burohman is also with the Bernoulli Institute for Mathematics, Computer Science and Artificial Intelligence, Faculty of Science and Engineering, University of Groningen, 9747AG Groningen, The Netherlands.}
\thanks{This work was supported by Islamic Development Bank.}}

\maketitle

\begin{abstract}
We propose a Kron-based model-order reduction method for 
mass-action kinetics chemical reaction networks (CRN) with constant inflow and proportional outflow. The reduced-order models preserve the CRN structure 
and we establish that the resulting reduced-order models have the same DC-gain or zero-moment as that of the full-order ones. Subsequently, we 
present the spectrum interlacing property of the Kron-reduced open CRN and propose the use of Gramians-based approach for single-species single-substrate chemical network to get the upper-bound of approximation error and to use it in determining a good set of nodes to be removed systematically. 
Finally, we evaluate the applicability and efficacy of our results in two well-known biochemical kinetic models: the activated sludge model (ASM) 1 and  McKeithan's T-cell receptor model.
\end{abstract}

\begin{IEEEkeywords}
Kron-based model-order reduction, moment matching, Gramians-based approach, mass-action kinetics, chemical reaction networks.
\end{IEEEkeywords}

\section{Introduction}

For the simulation and control of complex (bio)-chemical processes, kinetic models of the underlying chemical reaction networks are generally used. These models are given by ordinary differential equations whose order depends on the number of chemical species involved and their underlying kinetics. The resulting sets of ordinary differential equations for complex chemical processes are typically high dimensional (e.g., hundreds of species and reactions) and intrinsically nonlinear (e.g., polynomial if we assume the most basic form of modeling, e.g., mass action kinetics; or rational functions if we assume the standard Michaelis-Menten kinetics). 

The current state-of-the-art numerical tools for stability analysis, for bifurcation study (e.g., in \cite{Chickarmane2010}), for stochastic simulations and for other types of dynamical analysis are known to suffer from curse-of-dimensionality. Moreover since complex models of biochemical reaction networks involve a large number of parameters, the task of identifying these parameters (in addition to those parameters that have been identified in the literature) is enormous and requires large datasets. The complexity of this task is further compounded by the fact that often not all the species’ concentrations can be measured. Recently, there has also been an interest in kinetic modelling of genome-scale networks as presented in \cite{Smallbone2010} and \cite{Srinivasan2015}. Reduction of model size and complexity can thus help in focusing the computational analysis and control effort on important sub-species and sub-network. 

Using the chemical reaction networks (CRN) formulation as presented in \cite{AJS-16}, we can describe the dynamics of open CRN with inflow and outflow by
\begin{equation}\label{eq:open_CRN}
\dot x = ZDv(x) + ZD_{\text{in}}v_{\text{in}} + ZD_{\text{out}}v_{\text{out}}(x)
\end{equation}
where $Z$ is the complex\footnote{The notion of chemical complexes here refer to the combination of species that are involved in the left-hand and right-hand side of every reaction. The notion has played a key role in the mathematical analysis of chemical reaction networks in \cite{AJS-16} and \cite{FH-72}. It will be further discussed in Section \ref{sec:preliminary}.} composition matrix that maps chemical complexes to the individual chemical species, $D$, $D_{\text{in}}$ and $D_{\text{out}}$ are the incidence matrix of the internal network, of the inflow edges and of the outflow edges, respectively, and $x\in\mathbb{R}^n$ is the state space of chemical species. The reaction rate $v(x)\in\mathbb{R}^r$ is given by the underlying kinetics law of every reaction that can include the mass-action kinetics, Michaelis-Menten or other general kinetics law (see, for instance, \cite{BJ-12}). The influx rate $v_\text{in}\in\mathbb{R}^{r_\text{in}}$ and outflow $v_\text{out}(x)\in\mathbb{R}^{r_\text{out}}$ give the interaction of the 
CRN with the environment. In this formulation, the outflow rate can also be given by kinetics law and the inflow rate is typically given as a constant influx. For the genome-scale kinetics model, where the dimension of state space $n$ and reaction rate $r$ can reach tens of thousands, analyzing various stability and control properties of \eqref{eq:open_CRN} becomes a daunting task.

For closed chemical networks (without inflow and outflow), there are a number of model reduction techniques proposed in the literature. The time-scale separation technique as discussed in \cite{Radulescu2008}, \cite{Roussel2001}, \cite{Gorban2003}, \cite{Sunnaker2011} and the quasi steady-state approximation (QSSA) (see, for example, \cite{Segel1989}) are the most commonly used techniques. Hardin in \cite{Hardin2010} extends the QSSA approach by considering the higher-order approximation in the computation of the quasi steady-state. In \cite{Androulakis2000}, \cite{Bhattacharjee2003} and \cite{Petzold1999}, integer optimization techniques are used to determine which components and/or reactions can be eliminated from the original model without a substantial alteration of the model behaviour. Dan{\o} {\it et al} in \cite{Dano2006} propose a model reduction method by identification and elimination of variables in such a way that the basic dynamic properties of the original model are preserved. In \cite{Schmidt2008}, model reduction is achieved by simplifying the rate equations of individual enzymes in the network. In \cite{Apri2012}, it is rather the number of parameters of a given model that is reduced. This is done by identifying a region in the parameter space where some of the parameters can be pruned to zero while ensuring 
the outputs of the reduced model match those of the original model 
within a given tolerance. 

In recent years, a model reduction method of closed CRN using the graph information encoded in the incidence matrix $D$ in \eqref{eq:open_CRN} has been presented in \cite{SR-14}. The reduction of model order is achieved by pruning the network graph and by replacing it with an equivalent or approximated reduced network graph. This approach is motivated by the Kron reduction method used in electrical networks (see, e.g. \cite{GK-39}, \cite{AJS-10}, \cite{SYC-12}, and \cite{SYC-14}) and other complex networked systems, such as, \cite{FD-12}. The use of the Kron reduction method enables the preservation of the network graph structure and has been shown to be effective in reducing state equations in a number of well-known biochemical reaction networks as presented in \cite{SR-14} and in \cite{BJ-15}. Further work on this method can also be seen in \cite{MG-18} and \cite{MG-20}. There is also the method of structural reduction proposed by \cite{KMH-21} that can be applied to kinetic models with linear independent sub-CRNs consisting of linear reactions. Based on the detailed description of the linear sub-CRNs, the proposed method reduces them into a single reaction with distributed time delay. Therefore, one obtains a delayed CRN with conceivably different distributed time delays but with less complexes and reactions than the original model. 

On the other hand, for open chemical networks, which contain inflow and outflow as in \eqref{eq:open_CRN}, there are not many model reduction methods available that are specific for such open CRN. In \cite{ABSB-10}, it has been shown that a time-scale separation can be used to reduce the model via singular perturbation theory; akin to the QSSA approach for closed CRN above. In \cite{AP-22}, it introduced an automated model reduction pipeline based on QSSA where it gives robustness guarantees for structured model reduction of linear and nonlinear dynamical systems under parametric uncertainties. When the open CRN does not have a clear time-scale separation, this approach can no longer be applicable. In \cite{SR-14}, it is shown that the Kron reduction approach can be applicable to open CRN. Another work for open CRN 
is the use of balanced truncation applied to the associated variational systems obtained through linearization of the systems \cite{AP-21}. For open nonlinear CRN, it remains a challenge to solve the corresponding balanced truncation PDE for such networks \cite{JMS-93}. 

In this paper we present Kron-based model order reduction for open CRN following the approach in \cite{BJ-15} and we provide analysis on the mathematical properties that can be obtained by the Kron reduction approach. 
The first property corresponds to the zero-moment matching property of the Kron-based reduced-order open CRN. For linear systems, the zero-moment matching property corresponds to the preservation of the DC-gain in the reduced model. This notion is generalized to general nonlinear systems in \cite{AA-08} which zero-moment matching refers to the preservation of the input-output map when constant inputs are considered. As the steady-state property with constant inputs plays an important role in many applications of open CRN, the preservation of zero-moment by the model reduction methods is a highly desirable property, in addition to preserving the CRN structure and the chemical species. The second property is related to 
the spectrum property of the reduced open CRN model, such as, the eigenvalue interlacing property. 
The third property is on 
the use of Gramians-based approach to get the approximation error upper-bound when certain nodes are removed. This error upper-bound allows us to optimize the reduction network by determining the nodes to be removed with the smallest error upper-bound.

The rest of the paper is organized as follows. In Section \ref{sec:preliminary}, we present preliminaries and problem formulation. In Section \ref{sec:Kron}, Kron reduction method on open CRN is presented. In Section \ref{sec:Kron_moment}, we analyze the zero-moment matching property of the Kron reduction method. In Section \ref{sec:Gramians}, we present the use of Gramians-based approach to optimize the reduction network. In Section \ref{sec:example}, we show the performance of Kron reduction method in two different examples. Finally, the conclusions are given in Section \ref{sec:conclusion}.


\section{Preliminaries}\label{sec:preliminary}

\subsection{Graph formulation of CRN}\label{sec:graph_CRN}

Every chemical reaction network (CRN)  can be associated to a directed graph $\mathcal G=(\mathcal V,\mathcal E)$ with $\mathcal V$ be the set of vertices and $\mathcal E\subset \mathcal V\times \mathcal V$ be the set of edges, where every edge represents every reaction in CRN and every node in $\mathcal V$ corresponds to a chemical species (or a combination of chemical species) that is (or are) involved in the substrate (if the vertex is at the tail) or in the product of a reaction (if the vertex is at the head). The latter elements are usually referred to as the substrate and product complexes, respectively. Following the  graph-theoretical formulation of CRN in \cite{AJS-16}, we denote the species concentration by $x\in \rline^n$ with $n$ be the number of chemical species. We denote the number of chemical complexes by $c$ and that of reactions by $r$. The reaction edges in $\mathcal{G}$ can be represented by an incidence matrix $D\in\mathbb{R}^{c\times r}$ that maps a reaction to the constitutive substrate and product complexes. In other words, each column of $D$ corresponds to an edge of $\mathcal G$, and contains exactly one element of $1$ at the complex associated to the product (head vertex),  one element of $-1$ at the complex associated to the substrate (tail vertex) and all other elements are zero. The graph $\mathcal G$ is strongly connected if any vertex can be reached from any other vertex by a sequence of directed edges. A subgraph of $\mathcal{G}$ is a directed graph whose vertex and edge sets are subsets of the vertex and edge set of $\mathcal{G}$. For the CRN graph $\mathcal G$, we can relate the complexes with the underlying chemical species through the complex stoichiometric matrix $Z\in\mathbb{R}^{n\times c}$ where the numbers appeared in its $i$-th column represent the number of different species (associated to the row of $Z$) involved in the $i$-th complex. The usual stoichiometric matrix $S$ is given by $S=ZD$.

\subsection{Kron Reduction in Graph}

For a general connected undirected graph $\mathcal G$ (which can be interpreted as a strongly connected directed graph where each edge has the complement edge with different direction), the matrix $L = DD^T \in \mathbb{R}^{c\times c}$ is called the Laplacian matrix, and it is also known as graph Laplacian, admittance matrix or Kirchoff matrix in various different applications. The Laplacian matrix $L$ has been used to represent the graph of electrical circuits and power networks \cite{FD-12,AJS-10}, as well as, of balanced CRN with mass-action kinetics \cite{SR-13}. 
It satisfies   

$$0=\lambda_1(L)<\lambda_2(L)\leq...\leq\lambda_n(L).$$

For a given electrical circuit network $\mathcal G$, an equivalent electrical circuit sub-network can be obtained through the process known as the Kron reduction (see, e.g. \cite{SYC-14}). Specifically, if the associated Laplacian matrix $L$ can be decomposed into
\begin{equation}\label{eq:partition_L}
L=\begin{bmatrix}L_{11}&L_{12}\\L_{21}&L_{22}\end{bmatrix}.
\end{equation}
where $L_{11}$ is associated to the sub-graph of $\mathcal G$ with vertices $\mathcal V_{1} \subset \mathcal V$,  and $L_{22}$ represents the sub-graph of $\mathcal G$ with vertices $\mathcal V_2:=\mathcal V\backslash \mathcal V_{1}$ then the equivalent reduced circuit with $\mathcal V_{1}$ as its vertices has its Laplacian $\hat{L}$ given by 

\begin{equation*}
\hat{L}=L_{11}-L_{12}L_{22}^{-1}L_{21},
\end{equation*}
where $\hat{L}\in \mathbb{R}^{\hat c\times \hat c}$ with $\hat c = \dim{\mathcal V_{1}}$. As presented in \cite{FD-12},
the reduced Laplacian $\hat{L}$ is again a symmetric Laplacian matrix. Correspondingly one can obtain the reduced incidence matrix $\hat D$ so that $\hat{D}\hat{D}^T=\hat{L}$ which means that we can define the equivalent reduced edges. As discussed in \cite{FD-12} and \cite{BJ-15}, this Kron reduction process has the eigenvalues interlacing property where for every $i=1,\dots,\hat c$, 
\begin{equation}\label{eq:Laplacian_spectrum}
\begin{split}
\lambda_i(L)\leq\lambda_i(\hat{L})&\leq\lambda_{i+c-\hat c}(L)\end{split}\end{equation}
holds and it can be used to characterize the reduced network. 

\subsection{Moment Matching Model Reduction method}\label{subsec:moment_matching}

Let us briefly review the current literature on moment-matching model reduction methods in (non-)linear systems theory. We will later relate this moment-matching notion to the analysis of the model reduction method of CRN with in-flows and out-flows based on the Kron reduction approach. Consider the following nonlinear affine systems
\begin{equation}\label{eq:nonlin_SS}
\Sigma:  
\begin{array}{rl}
\dot{x}&=f(x)+g(x)u\\
y&=h(x),
\end{array}
\end{equation}
where $f,g,h$ are assumed to be smooth and $x(t)\in\rline^n, u(t),y(t)\in\rline$. The notion of moment-matching is closely related to the steady-state response of the system with respect to signals generated by an exosystem as presented in \cite{AA-07} and \cite{ACA-06}. In particular, it has been proposed in \cite{AA-07} that the reduced model of \eqref{eq:nonlin_SS} that matched with the moment generated by  
\begin{equation}\label{eq:exosystems}
    \dot w = s(w), \qquad u = c(w), 
\end{equation}
where $w(t)\in \Omega\subset \rline^m$ 
is given by the family of nonlinear systems 
\begin{equation}\label{eq:reduced_nonlin_SS}
\hat\Sigma:  
\begin{array}{rl}
\dot{\hat{x}}&=\hat{f}(\hat{x})+\hat{g}(\hat{x})u\\
y&=\hat{h}(\hat{x}),
\end{array}
\end{equation}
where $\hat{f},\hat{g},\hat{h}$ are smooth and the reduced state variable $\hat{x}(t)\in\rline^{\hat{n}}$ with $\hat n < n$, such that
\begin{equation}\label{eq:moment_matching_PDE}
\left. \begin{array}{rl} \frac{\partial \pi(w)}{\partial w}s(w) & = f(\pi(w)) + g(\pi(w))c(w) \\
\frac{\partial p(w)}{\partial w}s(w) & = \hat{f}(p(w)) + \hat{g}(p(w))c(w) \\
h(\pi(w)) & = \hat{h}(p(w))
\end{array} \right\}
\end{equation}
hold for all $w\in\Omega$, for some mappings $\pi:\rline^m\to\rline^n$ and $p:\rline^m\to\rline^{\hat{n}}$. In other words, the reduced-order systems \eqref{eq:reduced_nonlin_SS}  admit identical output trajectories with those of \eqref{eq:nonlin_SS} when they are subjected to the same input generated by \eqref{eq:exosystems}. In particular, when the exosystems \eqref{eq:exosystems} is given by an integrator with $s=0$, e.g. they are constant signal generators, \eqref{eq:moment_matching_PDE} becomes
\begin{equation}\label{eq:moment_matching_zero}
\left. \begin{array}{rl} 0 & = f(\pi(w)) + g(\pi(w))c(w) \\
0 & = \hat{f}(p(w)) + \hat{g}(p(w))c(w) \\
h(\pi(w)) & = \hat{h}(p(w)).
\end{array} \right\}
\end{equation}
When we restrict \eqref{eq:nonlin_SS} to the class of linear systems given by 
\begin{equation}\label{eq:linear_systems}
\Sigma:  
\begin{array}{rl}
\dot{x}&=Ax+Bu\\
y&=Cx
\end{array}
\qquad \text{ and } \qquad \hat{\Sigma}: 
\begin{array}{rl}
\dot{\hat{x}}&=\hat{A}\hat{x}+\hat{B}u\\
{\hat y}&=\hat{C}\hat{x},
\end{array}
\end{equation}
the conditions in \eqref{eq:moment_matching_zero} are equivalent to 
\begin{equation} \label{eq:zero_moment_matching}
CA^{-1}B=\hat{C} \hat{A}^{-1}\hat{B},
\end{equation}
which is a well-known condition for moment matching at zero frequency as presented in \cite{ACA-06}. In other words, the reduced-order models that match the moment at zero have the property that their DC gain matches that of the full-order ones. 

\subsection{Chemical Reaction Networks with in-/out-flows}

The model of a biochemical reaction network is a set of differential equations describing the evolution and dynamics of the concentrations of all the metabolites that are involved in the reaction network. This model involves some fixed parameters and some boundary fluxes which are usually functions of metabolite concentrations. The structure of a chemical reaction network cannot be directly captured by an ordinary graph. Instead, we will follow an approach in the work of \cite{AJS-16}.

The set of complexes of a chemical reaction network (CRN) is simply defined as the union of all the different left- and right-hand sides (substrate and product) complexes of the reactions in the network. The dynamics of a {\it closed} CRN can be given by 
\begin{equation}\label{eq:CRN}
\dot{x}=ZDv(x),
\end{equation}
where, as described before in subsection \ref{sec:graph_CRN}, $Z\in\rline_+^{n\times c}$ is the complex stoichiometric matrix of the network,  $D\in\rline^{c\times r}$ is the incidence matrix and 
$v(x)\in\mathbb{R}^r$ is the vector of reaction rates or fluxes. 
By defining $v(x)$ as a \emph{mass action kinetics}, as presented in \cite{AJS-16}, the reaction rate of the total reaction network is given by
\begin{equation}\label{eq:v(x)}
v(x)=K\text{Exp}(Z^T\text{Ln}x),
\end{equation}
where the \emph{outgoing co-incidence matrix}  $K
\in\rline^{r\times c}$ is the matrix whose $(j,\sigma)$th element equals the $j$-th reaction rate constant $k_j>0$ if the $\sigma$-th complex is the substrate complex for the $j$-th reaction. 
So that, the dynamics of mass action kinetics reaction takes the form
\begin{equation}\label{eq:complex-balanced_CRN}
\dot{x}=ZDK\text{Exp}(Z^T\text{Ln}x).
\end{equation}
It can be verified that the matrix $L:=-DK\in\rline^{c\times c}$ defines a weighted Laplacian matrix that has non-negative diagonal elements and non-positive off-diagonal elements. As stated in \cite{FH-72}, a CRN is called  \emph{complex-balanced} if there exists $x^*\in\rline^n$, called complex-balanced equilibrium,  satisfying
\begin{equation*}
Dv(x^*)=-L\text{Exp}(Z^T\text{Ln}x^*)=0.
\end{equation*}
This allow us to rewrite \eqref{eq:complex-balanced_CRN} into the form of
\begin{equation}\label{eq:CB_CRN}
\dot{x}=-Z\mathcal{L}(x^*)\text{Exp}\left(Z^T\text{Ln}\frac{x}{x^*}\right), \quad \mathcal{L}(x^*):=L\Xi(x^*),
\end{equation}
where $\Xi(x^*):=\text{diag}(\text{Exp}(Z^T\text{Ln}x^*))$, the operation $\frac{x}{x^*}$ is done in element-wise sense and  $\mathcal{L}(x^*)$ becomes a symmetric Laplacian matrix.

For a closed complex-balanced CRN 
with single-species single-substrate (which we will refer to as {\it SS reaction networks}) as studied in \cite{AJS-16}, we have 
$Z=I_c$ and the dynamics \eqref{eq:CRN} reduces to the following linear autonomous systems
\begin{equation}\label{eq:CRN_linear}
\dot{x}=DKx=-\mathcal{L}(x^*)\frac{x}{x^*} \text{ with } \mathcal{L}(x^*):=-DK\Xi(x^*).
\end{equation}

In general, biochemical reaction networks are not closed systems and they interact with the environment through additional inflow and outflow in some part of the network. Correspondingly, we can extend \eqref{eq:CRN} to an {\it open} CRN by incorporating these inflow and outflow as follows 
\begin{equation}\label{eq:CRN_i/o_base}
\left.\begin{array}{rl}
\dot{x}& =ZDv(x)+ZD_{\text{in}}v_{\text{in}}-ZD_{\text{out}}v_{\text{out}}(x)\\
y&=C\text{Exp}(Z^T\text{Ln}x), \end{array}\right\}
\end{equation}
where $D_{\text{in}}$ and $D_{\text{out}}$ are incidence matrices of the inflow and outflow that connect internal complexes to an additional ``zero''-complex $\emptyset$ in the vertex set $\mathcal V$ of CRN graph $\mathcal G$. In this formulation, the vector $v_{\text{in}}\in\mathbb{R}^c$ is the vector of inflow from the environment,  $v_{\text{out}}(x)\in\mathbb{R}^d$ is gives the outflow kinetics and $y$ is the measured output.  As before, we will assume throughout that the inflow $v_{\text{in}}$ are constant inflow and the outflow kinetics $v_{\text{out}}(x)$ are given by mass-action kinetics, e.g. 
\begin{equation}\label{eq:outflow_kinetics}
v_{\text{out}}(x)=K_{\text{out}}\text{Exp}(Z^T\text{Ln}x).
\end{equation}

In the context of moment-matching based model-order reduction, we consider $v_{\text{in}}$ as the input variable $u$ while the output variable $y$ will be the monitored chemical species which are part of $x$. For instance, when we again consider the SS reaction networks \eqref{eq:CRN_linear} with measured output variable $y=Cx$ for some selection matrix $C$ (comprising of only $1$ and $0$), the open SS CRN can be given by
\begin{equation}\label{eq:CRN_i/o_SS}
\left. \begin{array}{rl} 
\dot x & = \underbrace{-(\overbrace{DK}^{=:L}+\overbrace{D_{\text{out}}K_{\text{out}}}^{=:R})}_{=:A}x + \underbrace{D_{\text{in}}}_{=:B}u \\
y & = Cx \end{array} \right \}\end{equation}
where the matrices $A,B$ and $C$ are the usual matrices of linear systems as in \eqref{eq:linear_systems}.

Related to the zero-moment property as discussed in Subsection \ref{subsec:moment_matching}, it has been shown in \cite{AJS-16} on the existence of an attractive equilibrium manifold for a given constant inflow $v_{\text{in}}$ as follows.

\begin{proposition}[Theorem 4.4 in \cite{AJS-16}]
Consider a mass action kinetics reaction network with constant inflows and mass action kinetics outflows \eqref{eq:CRN_i/o_base}, for which there exists a complex-balanced steady state $x^*\in\mathbb{R}^n$ that satisfies
$Dv(x^*)+D_{\text{in}}v_{\text{in}}+D_{\text{out}}v_{\text{out}}(x^*)=0.$ 
Then 
\begin{itemize}
\item for every $x_0\in\mathbb{R}^n$, there exists a unique $x_1\in\mathcal{E}$ with $x_1-x_0\in\text{\normalfont\ im}ZD$ and $\mathcal{E}:=\{x^{**}\in \mathbb{R}^n|(ZD)^T\textup{Ln}(x^{**})=(ZD)^T\textup{Ln}(x^{*})\}$ be the equilibrium set; 
\item the steady state $x_1$ is locally asymptotically stable with respect to intial condition $x_0$; and
\item additionally, if the network is persistent\footnote{Following \cite{AJS-16}, a CRN is called persistent if no steady state can occur at the boundary of positive orthant $\rline^n_+$ whenever the initial states are all non-zero.} then $x_1$ is globally asymptotically stable with respect to all the initial conditions.
\end{itemize}
\end{proposition}

Based on this proposition, we will investigate in the rest of the paper on the property of $\mathcal E$ obtained from the original open CRN and that from the reduced-order one via Kron reduction approach. For the rest of the paper, we  consider complex-balanced CRN and we will study the preservation of steady-state input-output mapping $v_{\text{in}}\mapsto y$. 

\begin{figure}
\begin{center}
\begin{tikzpicture}[node distance={20mm}] 
\node (x1) {$x_1$}; 
\node (x2) [right of=x1] {$x_2$}; 
\node (x3) [right of=x2] {$x_3$}; 
\node (b) [above of=x1] {};
\node (a) [below of=x3] {};
\draw[->] (b) -- node[midway, right] {$k_{\textup{in}}$} (x1);
\draw[->] (x1) to[out=20,in=160] node[midway, above] {$k_1$} (x2); 
\draw[->] (x2) to[out=200,in=340] node[midway, below] {$k_{-1}$} (x1); 
\draw[->] (x2) to[out=20,in=160] node[midway, above] {$k_2$} (x3);
\draw[->] (x3) to[out=200,in=340] node[midway, below] {$k_{-2}$} (x2); 
\draw[->] (x3) -- node[midway, right] {$k_{\textup{out}}$} (a);
\end{tikzpicture}
\caption{Balanced chemical reaction network with inflow and outflow.} 
\label{fig:1}
\end{center}
\end{figure}
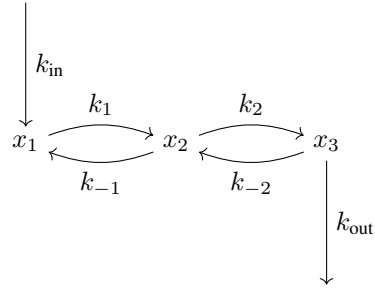

\begin{example}\label{ex_2_1}
Let us introduce a simple chemical reaction sub-network with inflow and outflow based on the well-known kinetic model of glycolysis studied in \cite{BT-00} and is shown in Fig.~\ref{fig:1}. We consider a sub-network of glycolysis kinetic model that involves the metabolic pathways of 3PGA, 2PGA and PEP while the influence from the other part of the network is considered as inflow and outflow. Fig.~\ref{fig:1} depicts this sub-network where $x_1,x_2$ and $x_3$ represent the metabolite concentrations of 3PGA, 2PGA and PEP, respectively. The constants $k_1,k_{-1},k_2,k_{-2}, k_{\text{in}}$ and $k_{\text{out}}$ in Fig. \ref{fig:1} are the rate constants, constant inflow and outflow rate constant, respectively. Following the rate constants used in \cite{BT-00} (excluding the nonlinear kinetic rate components that appear as the common denominator in each kinetic law in \cite{BT-00}), we will use $k_1=\frac{V_{\textup{max}}^\textup{PGM}}{K_m^\textup{PGM}}$, $k_{-1}=\frac{V_{\textup{max}}^\textup{PGM}}{K_m^\textup{PGM}\times K_{\textup{eq}}^\textup{PGM}}$, $k_2=\frac{V_{\textup{max}}^\textup{ENO}}{K_m^\textup{ENO}}$, $k_{-2}=\frac{V_{\textup{max}}^\textup{ENO}}{K_m^\textup{ENO}\times K_{\textup{eq}}^\textup{ENO}}$, $k_{\text{out}}=\frac{V_{\textup{max}}^\textup{PYK}}{K_m^\textup{PYK}}$ and $k_{\text{in}}=V_{\textup{max}}^\textup{PGK}$ throughout this paper. Using the numerical values in \cite{BT-00}, we have $k_1=7.83$, $k_{-1}=41.21$, $k_2=33.75$, $k_{-2}=5.04$, $k_{\text{out}}=7.64$ and $k_{\text{in}}=4.8$. However, as these constants do not satisfy the  Wegscheider's condition for detailed balanced CRN (see \cite{ANG-11}), e.g., they do not fulfill $k_1 k_2=k_{-1} k_{-2}$. Accordingly, we performed least square estimate to satisfy Wegscheider's condition with the above constants as the priors, which results into the following admissible parameters: 
$k_1=7.19$, $k_{-1}=41.11$, $k_2=32.53$, $k_{-2}=5.69$. Using these values, the kinetics of the glycolysis sub-network can be written as \eqref{eq:CRN_i/o_SS} with
\begin{equation}\label{eq:example_2_1}
\dot{x}=\underbrace{\bbm{-7.19&41.11&0\\7.19&-73.64&5.69\\0&32.53&-13.33}}_{A}x+\underbrace{\bbm{4.8\\0\\0}}_{B}u.
\end{equation}
In later sections we will refer again to this example and as a reference, we have here $L = \sbm{7.19&-41.11&0\\-7.19&73.64&-5.69\\0&-32.53&5.69}$ and $R=\sbm{0&0&0\\0&0&0\\0&0&7.64}$.
\end{example}

\subsection{Kron Reduction in Complex-Balanced Chemical Reaction Networks}

Let us revisit the Kron reduction method presented in \cite{SR-13} to reduce the kinetic model of chemical reaction networks. Consider the dynamics of a closed CRN with mass-action kinetics \eqref{eq:CB_CRN} as follows
\begin{equation*}
\dot{x}=Z\mathcal{L}(x^*)\text{Exp}(Z^T\text{Ln}\frac{x}{x^*}).
\end{equation*}
In order to apply Kron reduction, we partition the network where we split it to two sub-networks comprising of a sub-network that will be retained and another one that will be reduced. Correspondingly, let us partition $Z$ and $\mathcal L (x^*)$ into
\begin{equation}\label{eq:partition}
Z=\bbm{Z_1&Z_2} \, \text{and} \, \mathcal{L}(x^*)=\bbm{\mathcal{L}_{11}(x^*)&\mathcal{L}_{12}(x^*)\\\mathcal{L}_{21}(x^*)&\mathcal{L}_{22}(x^*)}.
\end{equation}
Following the approach in \cite{SR-13}, the corresponding reduced-order model is given by  
\begin{equation}\label{eq:CRN_red}
\hat{\Sigma}:\quad\dot{x}=-\hat{Z}\hat{\mathcal{L}}(x^*)\text{Exp}(\hat{Z}^T\text{Ln}\frac{x}{x^*}),
\end{equation}
where $\hat Z = Z_1$ and $\hat{\mathcal{L}}(x^*) = \mathcal{L}_{11}(x^*)-\mathcal{L}_{12}(x^*)\mathcal{L}_{22}^{-1}(x^*)\mathcal{L}_{21}(x^*)$. The  associated rows of $Z_2$ (corresponding to the removed complexes) that do not have commonalities in $Z_1$ give the subset of species $x$ that can be removed from the network.  

\section{Kron reduction method for open CRN with mass-action kinetics}\label{sec:Kron}

Let us consider an open CRN in \eqref{eq:CRN_i/o_base} with mass-action kinetics as in \eqref{eq:v(x)} and \eqref{eq:outflow_kinetics}. Since the outflow is given by \eqref{eq:outflow_kinetics}, the kinetics of open CRN can be written as 
\begin{equation}\label{eq:CRN_i/o}
\Sigma:\quad\dot{x}=-Z(L+R)\text{Exp}(Z^T\text{Ln}x)+ZD_{\text{in}}v_{\text{in}},
\end{equation}
where as in \eqref{eq:CRN_i/o_SS}, $R=D_{\text{out}}K_{\text{out}}$. 
Following the approach as before, let us use the partition $Z$ and $L$ as in \eqref{eq:partition_L} and let the incidence matrix of inflow $D_{\text{in}}$ and outflow rate constant $R$ be partitioned as follows  
\[
R=\bbm{R_{11} & 0\\ 0 & R_{22}} \text{ and }
D_{\text{in}}=\bbm{D_{\text{in},1} \\ D_{\text{in},2}}.
\]

For ease of expression, we consider the following auxiliary dynamical system
\begin{equation}\label{eq:aux_dyn}
\bbm{\dot{\xi}_1\\\dot{\xi}_2}\!=-\bbm{L_{11}+R_{11}&L_{12}\\L_{21}&L_{22}+R_{22}}\!\bbm{w_1\\w_2}+\bbm{D_{\text{in},1}v_{\text{in}}\\D_{\text{in},2}v_{\text{in}}},
\end{equation}
which corresponds to the dynamics of complexes in \eqref{eq:CRN_i/o} with $w_1=\text{Exp}(Z_1^T\text{Ln}x)$ and $w_2=\text{Exp}(Z_2^T\text{Ln}x)$. By imposing the constraint $\dot{\xi}_2=0$, it follows that
\begin{equation*}
w_2=-(L_{22}+R_{22})^{-1}(D_{\text{in},2}v_{\text{in}}-L_{21}w_1),
\end{equation*}
which can be substituted back to \eqref{eq:aux_dyn} leading to 
\begin{align*}
\dot{\xi}_1&=-((L_{11}+R_{11})-L_{12}(L_{22}+R_{22})^{-1}L_{21})w_1\\
&\qquad+(D_{\text{in},1}v_{\text{in}}-L_{12}(L_{22}+R_{22})^{-1}D_{\text{in},2}v_{\text{in}})
\end{align*}
By substituting $w_1=\text{Exp}(Z_1^T\text{Ln}x)$ and by considering the constrained equation $\dot{x}=\sbm{Z_1 & Z_2}\sbm{\dot{\xi}_1 \\ 0}=Z_1\dot{\xi}_1$, we obtain that the reduced network $\hat{\Sigma}$ is given by 
\begin{equation}\label{eq:CRN_i/o_red}
\hat{\Sigma}:\quad\dot{x}=-\hat{Z}\hat{L}\text{Exp}(\hat{Z}^T\text{Ln}x)+\hat{Z}\hat{D}_{\text{in}}v_{\text{in}},
\end{equation}
where $\hat Z = Z_1$, $\hat{L} = (L_{11}+R_{11})-L_{12}(L_{22}+R_{22})^{-1}L_{21}$ and $\hat{D}_{\text{in}}=D_{\text{in},1}-L_{12}(L_{22}+R_{22})^{-1}D_{\text{in},2}$. When $Z_2$ contains mappings from some species $x_i$ that do not appear in $Z_1$ then these species will be removed from the reduced model $\hat \Sigma$ in \eqref{eq:CRN_i/o_red}. 
For the output variable of reduced network $\hat \Sigma$, it is given by 
\begin{equation}\label{eq:output_reduced_network}
y=\hat{C}\text{Exp}(\hat{Z}^T\text{Ln}x),
\end{equation} 
where $\hat{C}=C_1-C_2(L_{22}+R_{22})^{-1}L_{21}$. 

Let us consider again the open SS CRN as in \eqref{eq:CRN_i/o_SS}. Since $Z=I$ in this case, the partitioning of $Z$, $L$, $R$ and $D_{\text{in}}$ as above corresponds to the partitioning of matrices $A, B$ and $C$ in \eqref{eq:CRN_i/o_SS} as follows
\begin{equation}\label{eq:Partition}A=\begin{bmatrix}A_{11}&A_{12}\\A_{21}&A_{22}\end{bmatrix},\quad B=\begin{bmatrix}B_1\\B_2\end{bmatrix}, \quad C=\bbm{C_1 & C_2}.\end{equation}
Hence the application of Kron reduction method to \eqref{eq:CRN_i/o_SS} gives 
\begin{equation}\label{eq:reduced_model_SS}
\left. \begin{array}{rl}
\dot{\hat x} & = \hat A\hat x + \hat B u \\
\hat y & = \hat C \hat x,
\end{array}
\right\}
\end{equation}
where 
$\hat{A}=A_{11}-A_{12}A_{22}^{-1}A_{21}$, 
$\hat{B}=B_1-A_{12}A_{22}^{-1}B_2$, and 
$\hat{C}=C_1-C_2A_{22}^{-1}A_{21}$.

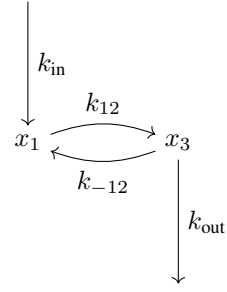
\begin{figure}
\begin{center}
\begin{tikzpicture}[node distance={20mm}] 
\node (x1) {$x_1$}; 
\node (x3) [right of=x1] {$x_3$}; 
\node (b) [above of=x1] {};
\node (a) [below of=x3] {};
\draw[->] (b) -- node[midway, right] {$k_{\textup{in}}$} (x1);
\draw[->] (x1) to[out=20,in=160] node[midway, above] {$k_{12}$} (x3); 
\draw[->] (x3) to[out=200,in=340] node[midway, below] {$k_{-12}$} (x1); 
\draw[->] (x3) -- node[midway, right] {$k_{\textup{out}}$} (a);
\end{tikzpicture}
\caption{Reduced balanced chemical reaction network with inflow and outflow.} 
\label{fig:3}
\end{center}
\end{figure}

\begin{example}[continues=ex_2_1]
Consider again the glycolysis metabolic sub-network example as shown in Fig.~\ref{fig:1} whose open CRN model is given by \eqref{eq:CRN_i/o_SS} with matrices $A$ and $B$ be as in \eqref{eq:example_2_1}. Suppose that the output matrix is given by $C=\bbm{0&0&1}$, e.g., we can measure the metabolite concentration of PEP. By using Kron reduction method, we can remove the metabolite concentration of $x_2$ (e.g., 2PGA) from the network and the resulting reduced open SS CRN is given by \eqref{eq:reduced_model_SS} with 
\begin{equation*}\hat{A}=\begin{bmatrix}3.18&-3.18\\-3.18&10.82\end{bmatrix},\quad
\hat{B}=\begin{bmatrix}4.8\\0\end{bmatrix},\text{ and } \hat{C}=\bbm{0&1}.\end{equation*}
The resulting reduced network is shown in Fig. \ref{fig:3} where the forward and reverse reaction rates are now given by $k_{12}=3.18=k_{-12}$ and $k_{\text{out}} = 7.64$ as before.
\end{example}

\section{Analysis of Kron-reduced open CRN}\label{sec:Kron_moment}

In the previous section, we have presented Kron reduction method that is applied to the open CRN kinetics in \eqref{eq:CRN_i/o}. In this section, we will investigate a number of systems' properties that can be preserved or obtained by the resulting Kron-reduced open CRN. 

\subsection{Zero-moment matching property}

For the steady-state or zero-moment matching property, we have the following affirmative result. 


\begin{proposition}\label{prop1}
Suppose that $\textup{Ker}(Z)=\emptyset$ and the underlying CRN graph $\mathcal G$ is undirected and connected. Then the zero-moment of reduced open CRN $\hat \Sigma$ in \eqref{eq:CRN_i/o_red} matches with the zero-moment of original open CRN $\Sigma$ in \eqref{eq:CRN_i/o}.
\end{proposition}

\begin{proof}
Consider the open CRN in 
\eqref{eq:CRN_i/o} with $A=L+R$ and $D_{\text{in}}$ with the corresponding output $y$ as follows
\begin{equation*}\nonumber
\Sigma:\left.\begin{array}{rl}\quad&\dot{x}=-ZA\text{Exp}(Z^T\text{Ln}x)+ZD_{\text{in}}v_{\text{in}},
\\&y=C\text{Exp}(Z^T\text{Ln}x). \end{array}\right.
\end{equation*}
where $Z$ is partitioned as in \eqref{eq:partition}, $A$, $D_{\text{in}}$ and $C$ are partitioned as in \eqref{eq:Partition}.

Now, let us analyze the zero-moment property of the original CRN, which satisfies
\begin{equation}\label{eq:zero_moment}
\left. \begin{array}{rl} &0=-ZA\text{Exp}(Z^T\text{Ln}x)+ZD_{
\text{in}}v_{\text{in}},
\\&y=C\text{Exp}(Z^T\text{Ln}x).
\end{array}
\right\}
\end{equation}
Since $Z$ has full column rank, the first equation in 
\eqref{eq:zero_moment} holds if and only if 
\begin{equation}\label{eq:zero_moment_1}
\left. \begin{array}{rl} &
0=-A\text{Exp}(Z^T\text{Ln}x)+D_{\text{in}}v_{\text{in}}.
\\&y=C\text{Exp}(Z^T\text{Ln}x).
\end{array}
\right\}
\end{equation}
Similarly, for the reduced-order open CRN in \eqref{eq:CRN_i/o_red}, its zero-moment satisfies
\begin{equation}\label{eq:zero_moment_red}
\left. \begin{array}{rl} &0=-Z_1(A_{11}-A_{12}A_{22}^{-1}A_{21})\text{Exp}(\hat{Z}^T\text{Ln}x) \\
& \qquad +  Z_1\Big(D_{\text{in},1}-(A_{12})(A_{22})^{-1}D_{\text{in},2}\Big) v_{\text{in}}
\\&y=(C_1-C_2(A_{22})^{-1}(A_{21}))\text{Exp}(\hat{Z}^T\text{Ln}x),
\end{array}
\right\}
\end{equation}
Since $Z$ is full column rank, we have that $Z_1$ is also full column rank. Hence the first equation in \eqref{eq:zero_moment_red} holds if and only if
\begin{equation}\label{eq:zero_moment_red_1}
\left. \begin{array}{rl} &0=-(A_{11}-A_{12}A_{22}^{-1}A_{21})\text{Exp}(Z^T\text{Ln}x) \\
& \qquad +  \Big(D_{\text{in},1}-(A_{12})(A_{22})^{-1}D_{\text{in},2}\Big) v_{\text{in}}
\\&y=(C_1-C_2(A_{22})^{-1}(A_{21}))\text{Exp}(\hat{Z}^T\text{Ln}x).
\end{array}
\right\}
\end{equation}

By the hypotheses of proposition, the matrix $A$ is invertible due to the connectedness of $\mathcal G$ and due to the fact that $R$ is a diagonal matrix with at least one positive entry (see, for example, Lemma 3 in \cite{WN-10}). Correspondingly, the first equation in \eqref{eq:zero_moment_1} satisfies $\text{Exp}(Z^T\text{Ln}x)=A^{-1}D_{\text{in}}v_{\text{in}}$ and by substituting this back to the second equation in \eqref{eq:zero_moment_1}, we obtain
\begin{equation}\label{eq:zero_moment_full}
y = CA^{-1}D_{\text{in}}v_{\text{in}}.
\end{equation}

For the Kron-reduced one in \eqref{eq:zero_moment_red_1}, we can have a similar expression as above. Firstly, we note that $A_{22}$ is invertible due to the connectedness of $\mathcal G$ so that its diagonal subblock is invertible\footnote{This can be shown by looking at the  sub-graph corresponding to the subblock elements, which has at least an outflow from an element in the sub-graph to another element in the rest of the subgraph. In this case, the invertibility of the subblock follows Lemma 3 in \cite{WN-10}.}. Consequently, it also follows that its Schur complement $(A_{11}-A_{12}A_{22}^{-1}A_{21})$ is invertible. This is due to the fact that $A$ is congruent with $\bbm{A_{11}-A_{12}A_{22}^{-1}A_{21} & 0 \\ 0 & A_{22}}$. Accordingly, by pre-multiplying \eqref{eq:zero_moment_red_1} with $(A_{11}-A_{12}A_{22}^{-1}A_{21})^{-1}$, we obtain that 
$\text{Exp}(\hat{Z}^T\text{Ln}x^*)=(A_{11}-A_{12}A_{22}^{-1}A_{21})^{-1}\hat{D}_{\text{in}}v_{\text{in}}$. Substituting this back to the second equation in \eqref{eq:zero_moment_red_1}, we get
\begin{equation}\label{eq:zero_moment_reduced}
y = \hat{C}(A_{11}-A_{12}A_{22}^{-1}A_{21})^{-1}\hat{D}_{\text{in}}v_{\text{in}},
\end{equation}
where $\hat{C}$ and $\hat{D}_{\text{in}}$ are as in \eqref{eq:CRN_i/o_red}. 

We will now show that \eqref{eq:zero_moment_full} is equivalent to \eqref{eq:zero_moment_reduced}. 
Using Schur complement (see \cite{JG-19}), we have that
\begingroup\scriptsize\begin{equation*} 
A^{-1} = \sbm{(A_{11}-A_{12}A_{22}^{-1}A_{21})^{-1}&-A_{11}^{-1}A_{12}(A_{22}-A_{21}A_{11}^{-1}A_{12})^{-1}\\-(A_{22}-A_{21}A_{11}^{-1}A_{12})^{-1}A_{21}A_{11}^{-1}&(A_{22}-A_{21}A_{11}^{-1}A_{12})^{-1}}.
\end{equation*}\endgroup
Hence for the relation in \eqref{eq:zero_moment_full}, we can have the value of $CA^{-1}D_{\text{in}}$. 
As before, since the graph $\mathcal G$ is undirected and connected, $A_{11}$ is invertible. Thus it follows that
\begingroup\small\begin{align*}
CA^{-1}D_{\text{in}}=&(C_1(A_{11}-A_{12}A_{22}^{-1}A_{21})^{-1}\\
&-C_2(A_{22}-A_{21}A_{11}^{-1}A_{12})^{-1}A_{21}A_{11}^{-1})D_{\text{in},1}\\
&-(C_1A_{11}^{-1}A_{12}(A_{22}-A_{21}A_{11}^{-1}A_{12})^{-1}\\
&-C_2(A_{22}-A_{21}A_{11}^{-1}A_{12})^{-1})D_{\text{in},2}.
\end{align*}\endgroup
By using matrix inversion lemma or Woodbury formula (see \cite{KSR-92}), it follows that \eqref{eq:zero_moment_full} is equivalent to \eqref{eq:zero_moment_reduced}.
\end{proof}

\begin{remark}
In Proposition \ref{prop1}, we assume that $\mathcal G$ is undirected and connected. This assumption can be weakened by having $\mathcal G$ directed and strongly connected. In this case, the claim of Proposition \ref{prop1} still holds if we assume additionally that the corresponding sub-block matrices $A_{11}$ and $A_{22}$ in the resulting leaky Laplacian matrix $A$ of $\Sigma$ are invertible.  
\end{remark}

For open SS CRN, the condition of  $\textup{Ker}(Z)=\emptyset$ in Proposition \ref{prop1} holds as $Z=I$.

\begin{example}
Let us consider again the biochemical reaction network in  Example \ref{ex_2_1} and the corresponding Kron-reduced CRN. From this numerical example, we can directly satisfy \eqref{eq:zero_moment_matching}. It can be computed that the zero-moment of both the full and reduced CRN satisfies $CA^{-1}B=\hat{C} \hat{A}^{-1}\hat{B}=-0.6283$. 
\end{example}


\begin{example}
Let us consider again the well-known kinetic model of glycolysis studied in \cite{BT-00} where we will take a sub-network with the following reactions  
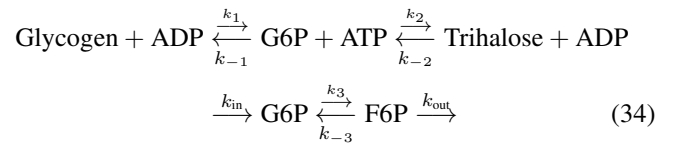
\begin{align}
\nonumber \textup{Glycogen}+\textup{ADP}&\overset{\overset{k_1}{\longrightarrow}}{\underset{k_{-1}}{\longleftarrow}}\textup{G6P}+\textup{ATP}\overset{\overset{k_2}{\longrightarrow}}{\underset{k_{-2}}{\longleftarrow}}\textup{Trihalose}+\textup{ADP}\\
\label{eq:nonlin_CRN} 
&\overset{k_{\text{in}}}{\longrightarrow} \textup{G6P}\overset{\overset{k_3}{\longrightarrow}}{\underset{k_{-3}}{\longleftarrow}}\textup{F6P}\overset{k_{\text{out}}}{\longrightarrow}
\end{align}
This sub-network involves the metabolic pathways of Glycogen, G6P, Trihalose, F6P, ADP and ATP while the influence from the other part of the network is considered as inflow and outflow to this sub-network. Let us denote $x_1=[\textup{Glycogen}],\; x_2=[\textup{G6P}],\; x_3=[\textup{Trihalose}],\; x_4=[\textup{F6P}],\; x_5=[\textup{ADP}]$ and $x_6=[\textup{ATP}]$ where $[X]$ denotes the concentration of $X$. 
The constants $k_1, k_{-1}, k_2, k_{-2}, k_3, k_{-3}, v_{\textup{in}}$ and $k_{\textup{out}}$ in \eqref{eq:nonlin_CRN} are the rate constants, constant inflow and constant outflow, respectively. Following the constants used in \cite{BT-00} and also using Wegscheider's condition, 
we consider $k_1=7.64,\; k_{-1}=6,\; k_2=2.4,\; k_{-2}=19.11,\; k_3=772.67,\; k_{-3}=242.62,\; k_{\textup{in}}=0.01$ and $k_{\textup{out}}=182.9$. Using these numerical values, the kinetics of the glycolysis sub-network can be written as
\begin{align*}
\dot{x}&=-ZA\sbm{x_1x_5\\x_2x_6\\x_2\\x_3x_5\\x_4}+ZD_{\textup{in}}v_{\textup{in}}\\
y&=C\sbm{x_1x_5\\x_2x_6\\x_2\\x_3x_5\\x_4}
\end{align*}
where
\begin{align*}
Z&=\sbm{1&0&0&0&0\\0&1&1&0&0\\0&0&0&1&0\\0&0&0&0&1\\1&0&0&1&0\\0&1&0&0&0},\; D_{\textup{in}}=\sbm{0\\0\\0.01\\0\\0},\; C=\sbm{0&0&1&0&0},\\
A&=\sbm{-7.64&6&0&0&0\\7.64&-8.4&0&19.11&0\\0&0&-772.67&0&242.62\\0&2.4&0&-19.11&0\\0&0&772.67&0&-425.52}.
\end{align*}
By using Kron reduction method, we can remove the last complex $x_4$ from the network and the resulting open CRN is given b
\begin{align*}
\dot{x}&=-\hat{Z}\hat{A}\sbm{x_1x_5\\x_2x_6\\x_2\\x_3x_5}+\hat{Z}\hat{D}_{\textup{in}}v_{\textup{in}}\\
y&=\hat{C}\sbm{x_1x_5\\x_2x_6\\x_2\\x_3x_5}
\end{align*}
where
\begin{align*}
\hat{Z}&=\sbm{1&0&0&0\\0&1&1&0\\0&0&0&1\\1&0&0&1\\0&1&0&0},\; \hat{D}_{\textup{in}}=\sbm{0\\0\\0.01\\0},\; \hat{C}&=\sbm{0&0&1&0},\\
\hat{A}&=\sbm{-7.64&6&0&0\\7.64&-8.4&0&19.11\\0&0&-332.11&0\\0&2.4&0&-19.11}.
\end{align*}
Since $Z$ and $\hat{Z}$ has full column rank, we can directly use \eqref{eq:zero_moment_full} and \eqref{eq:zero_moment_reduced} to calculate zero-moment for both the original network and the reduced network which is given by 
$3.011\times 10^{-4}$.
\end{example}

\subsection{Network spectrum interlacing property}

In \cite{AJS-13}, \cite{SR-13} and \cite{BJ-15}, it has been shown that the Kron reduction approach preserve the network structure of the original CRN. For instance, if the original CRN is detailed-balanced or complex-balanced then the Kron-reduced CRN is again detailed-balanced or complex-balanced, respectively. Another network property that is inherited by the Kron-reduced CRN is the network spectrum interlacing property where the spectrum of weighted Laplacian matrix of the Kron-reduced CRN is interlaced with that of the original CRN as in \eqref{eq:Laplacian_spectrum}. In this sub-section, we revisit this property again for the open CRN that contains inflow and outflow, in which case, the Laplacian matrix has an additional loss term of $R$.

\begin{proposition}\label{prop:interlacing}
For a given detailed-balanced open CRN as in \eqref{eq:CRN_i/o}, consider the corresponding Kron-reduced open CRN as in \eqref{eq:CRN_i/o_red}. Then $\sigma(\hat{L})$ interlace with $\sigma(L + R)$, i.e. for every $i=1,\ldots,\hat c$
\begin{equation}\label{eq:Kron_Spectrum}
\begin{split}0<\lambda_i&(L+R)\leq\lambda_i(\hat{L})\leq\lambda_{i+c-\hat c}(L+R),\end{split}\end{equation}
holds.
\end{proposition}
The proof of the proposition follows the standard result for Kron reduction of a positive semi-definite Hermitian matrix as in \cite{RLS-92} that is used for electrical networks in \cite{FD-12} or closed CRN in \cite{BJ-15}.


As presented in the preceding subsections, the outflow $R$ in open CRN plays a role 
in the reduced open CRN. The reduced CRN as given in  
\eqref{eq:CRN_i/o_red} shows that $R$ affects non-linearly to the expression of both $\hat{L}$ and $\hat{D}_{\text{in}}$. One can immediately notice from the structure of these matrices that if there is no outflow on the sub-graph that is removed from the network then we can have a direct relation between the reduced open CRN and the associated reduced {\em  closed} CRN. 

Indeed, suppose that $R_{22}=0$, i.e., there is no outflow from the removed complexes.  
Then $\hat{L}=\hat{L}_{\text{\rm closed}}+R_{11}$ where $\hat{L}_{\textrm{\rm closed}}$ is the Kron reduction of the weighted Laplacian matrix $L$ associated to the closed CRN. The absence of outflow $R_{22}$ has a direct 
effect also to the interlacing properties of the detailed-balanced CRN and its reduced network.



\begin{corollary}
Consider a given detailed-balanced open CRN as in \eqref{eq:CRN_i/o} and its reduced network as in \eqref{eq:CRN_i/o_red}. If $R_{22}=0$ 
then $\sigma(\hat{L}_{\text{\rm closed}}+{R}_{11})$, where  $\hat{L}_{\text{\rm closed}}=L_{11}-L_{12}L_{22}^{-1}L_{21}$, interlaces with $\sigma(L + R)$, i.e. 
\begin{equation}\label{eq:Kron_Spectrum_ex}
\begin{split}0<\lambda_i&(L+R)\leq\lambda_i(\hat{L}_{\text{\rm closed}}+{R}_{11})\leq\lambda_{i+c-\hat c}(L+R)\end{split}\end{equation}
holds for all $i=1,\ldots,\hat c$.
\end{corollary}

\begin{example}
Let us consider again the biochemical reaction network in  Example \ref{ex_2_1} and the corresponding Kron-reduced open CRN. For this numerical example, we have that $c=3$ and $\hat c = 2$, in which case, the application of Proposition \ref{prop:interlacing} gives the following relationship
\begin{align*}
\lambda_1(L+R)& \leq \lambda_1\left(\hat{L}_{\text{\rm closed}}+R_{11}\right)  \\
& \leq \lambda_2(L+R)\leq \lambda_2\left(\hat{L}_{\text{\rm closed}}+R_{11}\right) \leq \lambda_3(L+R),
\end{align*}
where $R_{11}=\sbm{0&0\\0& 7.64}$. 
Indeed, direct computation of the eigenvalues of $L+R$ and $\hat{L}_{\text{\rm closed}}$ shows that
$\lambda_1(L+R)=1.8745$, $\lambda_2(L+R)=11.8516$,  $\lambda_3(L+R)=80.4339$, 
$\lambda_1\left(\hat{L}_{\text{\rm closed}}+R_{11}\right)=2.0281$, and  $\lambda_2\left(\hat{L}_{\text{\rm closed}}+R_{11}\right)=11.9645$. 


\end{example}

\section{Selection of removed complexes of open SS CRN via Generalized Gramians}\label{sec:Gramians}
As shown in \cite{SR-14}, the selection of removed nodes using Kron reduction method in a closed CRN plays an important role in the quality of the approximation error. Correspondingly, Rao {\it et al.} has proposed the combined use of error integral and simulation in \cite{SR-14} to remove one node at a time, in order to obtain the set of removed nodes. In this section, we will propose the use of generalized Gramian 
to get the optimal set of removed nodes along with the model reduction error bound for a class of open detailed-balanced SS CRN. 

\subsection{Generalized Gramians}
Let us consider again the linear systems and their reduced ones as in \eqref{eq:linear_systems}. For the linear systems $\Sigma$, controllability and observability Gramians have been used to obtain the reduced-order models $\hat\Sigma$, see, e.g. \cite{antoulas2005approximation}. These Gramians reveal the states of systems that are hard to control and observe. Instead of using the ordinary Gramian to get the controllability or observability Gramian, generalized Gramians can be defined to characterize state variables that are difficult to control or to observe. In particular, {\it generalized controllability Gramians} is defined as a solution of inequality
\begin{equation}\label{eq:Gctrbgram}
    AP+PA^T+BB^T \leq 0,
\end{equation}
and, similarly, {\it generalized observability Gramians} is a solution of inequality
\begin{equation}\label{eq:Gobsvgram}
    A^T Q+QA+C^TC \leq  0.
\end{equation}
Note that, the matrices $P$ and $Q$ in \eqref{eq:Gctrbgram} and \eqref{eq:Gobsvgram} are not unique and satisfy $P\geq P_0$ and $Q \geq Q_0$ with $P_0$ and $Q_0$ be the usual controllability and observability Gramian, respectively. This non-uniqueness gives extra degree of freedom on their structure. Namely, we can force $P$ and $Q$ to have a specific structure, such as forcing $P$ and $Q$ to be diagonal. Balanced truncation method can also be applied by using this generalized Gramians, where $P$ and $Q$ are treated similarly as ordinary Gramians $P_0$ and $Q_0$ (see, for instance, \cite{dullerud2013course}).

\subsection{Generalized Gramians of complex-balanced SS CRN}
Since $P$ and $Q$ are not unique, 
we can directly compute diagonal $P$ and $Q$ such that \eqref{eq:Gctrbgram} and \eqref{eq:Gobsvgram} are satisfied. 
While the quantities in the diagonal of $P$ and $Q$ from \eqref{eq:Gctrbgram} and \eqref{eq:Gobsvgram} are not necessarily ordered, 
they give information on the states that are hard to control and to observe. 
Based on this information, one can 
select which nodes to be clustered in a clustering-based model reduction of networked systems as pursued in \cite{besselink2015clustering}, which is related to 
Kron reduction method applied to edge dynamics of networked systems. Motivated by the results in 
\cite{besselink2015clustering}, we will use generalized Gramians to provide a systematic method to determine a set of complexes of open SS CRN that can be removed via Kron reduction. This approach is in contrast to the approach in \cite{SR-14}, where one complex is removed at each reduction step instead of finding a set of complexes simultaneously. Correspondingly, we will focus 
on the reduction of open SS CRN 
which takes the form of linear systems as in  \eqref{eq:CRN_i/o_SS}. We formalize the generalized Gramians in the following definition. 

\begin{definition}\label{def:gen_gram}
Matrices $P\in \rline_+^{n\times n}$ and $Q\in\rline_+^{n\times n}$ are said to be \emph{generalized controllability} and \emph{observability Gramians of open SS CRN systems} \eqref{eq:CRN_i/o_SS} if they are diagonal and satisfy 
\begin{equation}\label{eq:def_ctrbgram}
        AP+PA^T+BB^T \leq 0
\end{equation}
and 
\begin{equation}\label{eq:def_obsvgram}
        A^TQ+QA+A^TC^TCA \leq 0,
\end{equation}
respectively, where $A$, $B$ and $C$ are as in \eqref{eq:CRN_i/o_SS}. 
\end{definition}
We remark that the matrix inequality \eqref{eq:def_obsvgram} is stronger than the one defined in \eqref{eq:Gobsvgram}. It can be verified that if 
$Q^*$ is a solution of \eqref{eq:Gobsvgram} then $A^TQ^*A$ is a solution of \eqref{eq:def_obsvgram}. 
The generalized Gramians of open SS CRN in  Definition~\ref{def:gen_gram} will allow for 
the computation of error bounds in Proposition~\ref{thm:upper_bound} below. In this regards, the computation of tight model reduction error bounds via 
\eqref{eq:def_ctrbgram} and \eqref{eq:def_obsvgram} can be done by 
minimizing $\mathrm{trace}(P)$ and $\mathrm{trace}(Q)$. 

Following Definition~\ref{def:gen_gram}, we can express the generalized Gramians as
\begin{equation}\label{eq:gen_Gramians}
    P= \begin{bmatrix}
    \pi^c_1 & & \\
     & \ddots & \\
     & & \pi^c_n
    \end{bmatrix}
    \quad \text{and} \quad  Q= \begin{bmatrix}
    \pi^o_1 & & \\
     & \ddots & \\
     & & \pi^o_n
    \end{bmatrix}.
\end{equation}
Before we further discuss about Kron reduction with respect to a subset of complexes, 
let us first consider a one step Kron reduction, where we only remove one complex that is deemed the least controllable and observable from generalized Gramian standpoint, as follows. 
\begin{proposition}\label{prop:Gram_red}
Consider an open SS CRN system $\Sigma$ as in \eqref{eq:CRN_i/o_SS} and its reduced-order model $\hat{\Sigma}_{n-1}$ via Kron reduction as in \eqref{eq:reduced_model_SS} by removing the $n$-th node so that 
the reduced system 
$\hat{\Sigma}_{n-1}$ are given by system matrices $\hat{A}=A_{11}-A_{12}A_{22}^{-1}A_{21}$, $\hat{B}=B_1-A_{12}A_{22}B_2$ and $\hat{C}=C_1$. Then 
\begingroup\small\begin{equation}
    \hat{P}_1= \begin{bmatrix}
    \pi^c_1 & & \\
     & \ddots & \\
     & & \pi^c_{n-1}
    \end{bmatrix}
    \quad \text{and} \quad  \hat{Q}_1= \begin{bmatrix}
    \pi^o_1 & & \\
     & \ddots & \\
     & & \pi^o_{n-1}
    \end{bmatrix},
\end{equation}\endgroup
are generalized controllability and observability Gramians for system $\hat{\Sigma}_{n-1}$, respectively.
\end{proposition}
\begin{proof}
By considering a one-step reduction, i.e., truncating only a single node, the partition of matrix $A$ is given by \eqref{eq:Partition}
with $\hat{A} \in \mathbb{R}^{(n-1) \times (n-1)}$ and $A_{22}$ is a scalar. Applying projection matrix $T_c= \begin{bmatrix} I & -A_{12}A_{22}^{-1} \end{bmatrix}$, we obtain
\begin{equation}\label{eq:lyap_ineq_ctrb}
T_c(AP+PA+BB^T)T_c^T = \hat{A} \hat{P}_1 + \hat{P}_1 \hat{A} + \hat{B} \hat{B}^T \leq 0.
\end{equation}
It is clear from the right hand side of \eqref{eq:lyap_ineq_ctrb} that the $\hat{P}_1$ is a generalized controllability Gramian for system $\hat{\Sigma}_{n-1}$. The proof for the other item is similar. Namely, applying projection $T_o=\bbm{ I & -A_{21}^T A_{22}^{-T}}$, we obtain
\begingroup\small\begin{equation*}
    T_o (A^T Q + Q A + A^T C^T C A) T_o^T = \hat{A}^T \hat{Q}_1 + \hat{Q}_1 \hat{A} + \hat{A} C_1^T C_1 \hat{A} \leq 0,
\end{equation*}\endgroup
which shows us that $\hat{Q}_1$ is a generalized observability Gramian of system $\hat{\Sigma}_{n-1}$.
\end{proof}

Note that in Proposition~\ref{prop:Gram_red}, it is assumed that $\hat{C}=C_1$. Namely, $C_2$ is assumed to be zero. 
This corresponds to the situation when we will only remove  complex(es) that is not measured directly. This assumption is reasonable in this CRN model as the measured species from the reduced-order model and the original model is supposed to be preserved and coincide. We will use this assumption for the rest of this section.

In the next result, we will show that if we consider detailed-balanced SS CRN system, then an a priori upper bound can be obtained using 
the generalized Gramians. 
\begin{proposition}\label{thm:upper_bound}
Consider an open SS CRN system $\Sigma$ and the corresponding one-step Kron reduced-order model 
$\hat{\Sigma}_{n-1}$ as assumed in Proposition \ref{prop:Gram_red}. 
Then for any input function $u(\cdot) \in \mathcal{L}_2[0,\infty)$ and initial condition $x(0)=0$ and $\hat{x}(0)=0$, the outputs satisfy
\begin{equation}\label{eq:bound_y/u}
    \lVert y - \hat{y} \rVert_2 \leq 2 M_{22} \sqrt{\left(\pi_n^c \pi_n^o \right)} \lVert u \rVert_2,
\end{equation}
where the scalar $M_{22}>0$ is a diagonal element of the partition matrix
\begin{equation}\label{eq:partition_M}
    M := \begin{bmatrix}
    M_{11} & M_{12} \\ M_{21} & M_{22} \end{bmatrix} = -A^{-1}=(L+R)^{-1},
\end{equation}
and $\pi_{n}^c,$ $\pi_{n}^o$ are the corresponding removed elements 
of the generalized Gramians $P$ and $Q$ as in \eqref{eq:gen_Gramians}. 
\end{proposition}
\begin{proof}
The proof of proposition is based on that of \cite[Theorem~11]{besselink2015clustering}. Let us analyze the error between systems $\Sigma$ and $\hat{\Sigma}_{n-1}$. For facilitating the analysis, 
instead of analyzing 
\eqref{eq:CRN_i/o_SS}, we will consider another state-space representation 
by pre- multiplying \eqref{eq:CRN_i/o_SS} with $M=-A^{-1}$:
\begin{equation}\label{eq:sys_ori_pf}
    M \dot{x} = - I x + MB u, \quad y=Cx.
\end{equation}
Before defining its 
reduced-order system, 
we note that 
$M$ can be partitioned as follows 
\begingroup\small\begin{equation}
    \bbm{M_{11} & M_{12} \\ M_{21} & M_{22}}= \bbm{M_{11} & -M_{11}A_{12}A_{22}^{-1} \\ -M_{22}^{-1}A_{21}A_{11}^{-1} & M_{22}},
\end{equation}\endgroup
where $M_{11}=-(A_{11}-A_{12}A_{22}^{-1}A_{21})^{-1}=-\hat{A}^{-1}$. 
Correspondingly, 
the reduced-order model $\hat{\Sigma}_{n-1}$ in \eqref{eq:reduced_model_SS} can also be rewritten 
by pre-multiplying it by $M_{11}$, which gives us  
\begin{equation}\label{eq:sys_red_pf}
    M_{11} \dot{\hat{x}} = -I \hat{x} + M_{11}\hat{B}u, \quad \hat{y}=C_1 \hat{x},
\end{equation}
where $\hat{B}=B_1-A_{12}A_{22}^{-1}B_2$ and we have used the assumption that $C_2=0$. 

In the remainder of this proof, we analyze the error between systems \eqref{eq:sys_ori_pf} and \eqref{eq:sys_red_pf} via frequency-domain analysis. 

Let the transfer function of \eqref{eq:sys_ori_pf} and \eqref{eq:sys_red_pf} be denoted by $G(s)$ and $\hat{G}(s)$, respectively. In order to write the transfer function error $G(s)-\hat{G}(s)$ in a convenient form, we first express the inverse of the partitioned matrix using Schur's complement as follows 
\begingroup\footnotesize\begin{align}
\begin{split}
    (sM+I)^{-1} = &\bbm{\varphi(s) & 0\\0 & 0} \\
    &+\bbm{-s\varphi(s)M_{12} \\ I}\Delta^{-1}(s) \bbm{-M_{21}s\varphi(s) & I},
    \end{split}
\end{align}\endgroup
where
\begin{align}
\begin{split}
    \varphi(s) &= (sM_{11}+I)^{-1},\\
    \Delta(s) &= sM_{22}+1-s^2M_{21} \varphi(s) M_{12},
    \end{split}
\end{align}
and $M_{22}$ is a scalar. Therefore, we can write the transfer function error as $G(s)-\hat{G}(s)=\tilde{C}(s) \Delta^{-1}(s) \tilde{B}(s)$ where 
\begin{align}
    \tilde{C}(s) &= {C}_1s \varphi(s) M_{12}  \label{eq:C(s)}\\
    \tilde{B}(s) &= \bbm{M_{21}s\varphi(s) & -I}\bbm{M_{11} & M_{12}\\M_{21} & M_{22}} \bbm{B_1\\ B_2}.
\end{align}
Let us quantify the error using the following $\cal{H}_{\infty}$ norm 
\begin{equation}\label{eq:sigmamax}
\lVert G(s)-\hat{G}(s)\rVert_{\cal{H}_{\infty}} =
\sup_{\omega \in \mathbb{R}} \sigma_{\max} 
\left(G(j\omega)- \hat{G}(j\omega)\right),
\end{equation}
where $\sigma_{\max}(\cdot)$ denotes the largest singular value. Using the fact that $\Delta(s)$ and the product $\tilde{B}(s)\tilde{B}(s)^T$ are scalar, it follows that 
\begingroup\footnotesize\begin{equation}
    \sigma_{\max}^2 \left( G(j\omega)-\hat{G}(j\omega) \right)= \dfrac{\tilde{B}(j\omega)\tilde{B}^H(j \omega)}{\Delta(j\omega) \Delta^H(j \omega)} \lambda_{\max} \left(\tilde{C}(j \omega)\tilde{C}^H(j\omega)\right),
\end{equation}\endgroup
where $G^H$ denotes the Hermitian transpose satisfying $G^H(j\omega) = (G(-j\omega))^T$ and $\lambda_{\max}$ denotes the largest eigenvalue. From the structure of $\tilde{C}(s)$ in \eqref{eq:C(s)}, it is clear that $\tilde{C}(j \omega)\tilde{C}^H(j\omega)$ is of rank one and therefore, the maximum eigenvalues is the only non-zero eigenvalue. Namely, $\lambda_{\max} \left(\tilde{C}(j \omega)\tilde{C}^H(j\omega)\right)=\tilde{C}^H(j \omega)\tilde{C}(j\omega)$, which leads to 
\begingroup\footnotesize\begin{equation}\label{eq:sigma_scalar}
    \sigma_{\max}^2 \left( G(j\omega)-\hat{G}(j\omega) \right)= \dfrac{\tilde{B}(j\omega)\tilde{B}^H(j \omega) \tilde{C}^H(j \omega)\tilde{C}(j\omega)}{\Delta(j\omega) \Delta^H(j \omega)} .
\end{equation}\endgroup

In order to provide an upper bound of \eqref{eq:sigma_scalar}, we will consider terms $\tilde{B}(j\omega)\tilde{B}^H(j \omega)$, $\tilde{C}^H(j \omega)\tilde{C}(j\omega)$ and $\Delta(j\omega) \Delta^H(j \omega)$ separately. As in the proof of \cite[Theorem~11]{besselink2015clustering}, by exploiting the matrix inequalities of the controllability Gramian \eqref{eq:def_ctrbgram}, we have the bound 
\begin{equation}\label{eq:boundB}
    \tilde{B}(j\omega)\tilde{B}^H(j \omega) \leq N(j \omega)\pi_n^c + \pi_n^c N^H(j\omega),
\end{equation}
where 
\begin{equation}
    N(j \omega) = M_{22}-M_{21}j \omega \varphi(j\omega) M_{12}.
\end{equation}
Similarly, by using matrix inequality of the observability Gramian \eqref{eq:def_obsvgram}, we obtain the bound
\begin{equation}\label{eq:boundC}
    \tilde{C}^H(j \omega)\tilde{C}(j\omega) \leq N^H(j \omega)\pi_n^o + \pi_n^o N(j\omega).
\end{equation}
By collecting the bounds \eqref{eq:boundB} and \eqref{eq:boundC}, we have the bound
\begin{equation}\label{eq:sigma_max_G-G}
    \sigma_{\max}^2 \left( G(j\omega)-\hat{G}(j\omega) \right) \leq \pi_n^c \pi_n^o \delta(\omega),
\end{equation}
where the real-valued function $\delta(\omega)$ is defined as 
\begin{equation}\label{eq:delta_w}
    \delta(\omega) = \dfrac{\Big(N(j\omega)+N^H(j\omega)\Big)^2}{\Delta(j\omega) \Delta^H(j\omega)}.
\end{equation}
In the remainder of the proof, we will show that $\delta$ satisfies
\begin{equation}\label{eq:sup_delta}
    \sup_{\omega \in \mathbb{R}} \delta(\omega) = \delta(0)=4M_{22}^2.
\end{equation}
Before we prove the upper bound of $\delta$, let us briefly review a number of important facts. 
We first note that 
$M>0$, which implies that 
$M_{22}>0$ since it is the inverse of `leaky' Laplacian matrix $L+R$. Additionally, since 
we consider detailed-balanced systems, 
there exist a diagonal matrix $\Xi(x^*)$ such that $A\Xi(x^*)=-(L+R)\Xi(x^*)$ is a symmetric matrix. Let us partition $\Xi$ into 
\begin{equation}
    \Xi(x^*) = \bbm{\Xi_1 & \\ & \Xi_2},
\end{equation}
with a scalar $\Xi_2$.
Correspondingly, by using the partition of matrix $A$ as in \eqref{eq:Partition}, we also have that $\hat{A}\Xi_1$ is also a symmetric matrix and 
admits eigenvalue decomposition $\hat{A}\Xi_1=U \Lambda U^T$. Thus 
\begin{equation}\label{eq:decom_A}
    \hat{A}=U \Lambda U^T \Xi_1^{-1},
\end{equation}
where $\Lambda$ is a diagonal matrix of the eigenvalues of $\hat{A}\Xi(x^*)$ and $U$ is an orthogonal matrix. The last fact corresponds to the 
symmetric matrix $A\Xi(x^*)$ which implies that 
\begin{equation}\label{eq:sym_A12}
    A_{12}\Xi_2 = \Xi_1 A_{21}^T,
\end{equation}
where $\Xi_2$ is a scalar. After reviewing these facts, let us 
return to the matrices in \eqref{eq:delta_w}. The terms $N(j\omega)$ can be written as
\begingroup\footnotesize\begin{align}\label{eq:nabla_A}
\begin{split}
N(j\omega) = (M_{22}-M_{21}M_{11}^{-1}M_{12})
 + M_{21}M_{11}^{-1}(j\omega I-\hat{A})^{-1}M_{11}^{-1}M_{12}.
\end{split}
\end{align}\endgroup
Using the eigenvalue decomposition \eqref{eq:decom_A} in \eqref{eq:nabla_A}, we obtain
\begin{equation}\label{eq:nabla_w/c}
    N(j\omega)= (M_{22}-M_{21}M_{11}^{-1}M_{12}) + \sum_{i=1}^{n-1} \dfrac{c_i}{j\omega+\lambda_i},
\end{equation}
where 
\begin{equation}
    c_i = M_{21}M_{11}^{-1}U_iU_i^T \Xi_1^{-1}M_{11}^{-1}M_{12},
\end{equation}
 $U_i$ denotes the $i$-th column of $U$ and $\lambda_i$ is the corresponding positive eigenvalue. From the partition of matrix $M$, we have that $M_{21}M_{11}^{-1}=-A_{22}^{-1}A_{21}$ and $M_{11}^{-1}M_{12}=-A_{12}A_{22}^{-1}$. This leads to \begin{equation}
     c_i = A_{22}^{-1}A_{21}U_iU_i^T\Xi_1^{-1}A_{12}A_{22}^{-1}.
 \end{equation}
 By noting that $A_{22}$ and $\Xi_2$ are scalar, it follows from 
 \eqref{eq:sym_A12} and the above relation that 
 \begin{equation}
     c_i =  \left\lvert \Xi_2^\frac{1}{2} U_i^T A_{21}^T A_{22}^{-1} \right\rvert^2,
 \end{equation}
which shows that $c_i\geq 0$. 

With regards to the numerator of \eqref{eq:delta_w}, we have that 
\begingroup\small\begin{equation}\label{eq:nabla+nabla}
    N(j \omega) + N^H(j\omega) = 2(M_{22}-M_{21}M_{11}^{-1}M_{12})+\sum_{i=1}^{n-1} \dfrac{2c_i \lambda_i}{\omega^2+\lambda_i^2}.
\end{equation}\endgroup
From the fact that $c_i\geq 0$ and $\lambda_i >0$, it follows that \eqref{eq:nabla+nabla} decreases as $\omega$ increases. Therefore, it is clear that 
\begin{equation}
    \sup_{\omega \in \mathbb{R}} \left(N(j \omega) + N^H(j\omega) \right)= N(0)+N^H(0) =2M_{22}.
\end{equation}
Now, we consider the denominator of \eqref{eq:delta_w}. It can be shown that $\inf_{\omega \in \mathbb{R}} \Delta(j\omega)\Delta^H(j\omega)=\Delta(0)\Delta^H(0)=1$. Consequently,
\begin{equation}\label{eq:sup_delta_w}
    \sup_{\omega \in \mathbb{R}} \delta(\omega) \leq \dfrac{\sup_{\omega \in \mathbb{R}} \left(N(j \omega) + N^H(j\omega) \right)^2}{\inf_{\omega \in \mathbb{R}} \Delta(j\omega)\Delta^H(j\omega)}=4M_{22}^2. 
\end{equation}
By substituting \eqref{eq:sup_delta_w} in \eqref{eq:sigma_max_G-G}, we obtain 
\begin{equation}
    \sigma_{\max}\left( G(j\omega)-\hat{G}(j\omega) \right) \leq 2M_{22} \sqrt{\pi_n^c \pi_n^o},
\end{equation}
which is equivalent to \eqref{eq:bound_y/u}.
\end{proof}
\begin{remark}\label{re:detailed_balanced}
In the proof of Proposition \ref{thm:upper_bound} above, the detailed-balanced assumption is used to guarantee the positive semi-definiteness of $c_i$. If the hypothesis is relaxed to 
complex balanced CRN, 
it remains an open problem whether 
$c_i \geq 0$ can be guaranteed. 
However, if we can guarantee that \eqref{eq:sup_delta_w} holds for a given partition matrix $M$, then the bound \eqref{eq:bound_y/u} also holds for general complex-balanced systems. 
Hence, we can use the upper bound \eqref{eq:bound_y/u} for non detailed-balanced systems as will be applied to a numerical example in Section~\ref{sec:example}.
\end{remark}

At this point, we have not discussed the procedure to obtain optimal choice of nodes to be removed. Accordingly, we can use 
the obtained upper bound in  Proposition \ref{thm:upper_bound}.
In particular, based on the result in 
Proposition \ref{thm:upper_bound}, we can order the complexes (or vertices of CRN) 
such that 
\begin{equation}\label{eq:numbering}
    M_{11}^2 \pi_1^c\pi_1^o \geq \dots \geq M_{nn}^2 \pi_n^c\pi_n^o \geq 0.
\end{equation}
Based on this order, we can consider the removal of complexes associated to smallest error bound. 
By removing the vertex corresponding to the smallest error bound, we can guarantee that the reduced-order model will have a small approximation error, but not necessarily the smallest. 
Note that such ordering procedure corresponds simply to 
applying a coordinate transformation $Tx$ using a permutation matrix $T$. 

\begin{example}\label{ex:gramian_based}
Let us consider again the biochemical reaction network in  Example \ref{ex_2_1} and the corresponding Kron-reduced CRN. For this numerical example, we have
\begin{equation}
    M= \bbm{
    0.4458  &  0.3067 &   0.1309 \\
    0.0536  &  0.0536 &   0.0229 \\
    0.1309 &   0.1309 &   0.1309
    }.
\end{equation}
By solving the matrix inequalities \eqref{eq:def_ctrbgram} and \eqref{eq:def_obsvgram}, we obtain
\begin{align}
    P&=\mathrm{diag}(6.1949,0.6885,2.1055),\\
    Q&=\mathrm{diag}( 2.7773,16.3089,10.0080).
\end{align}
These matrices lead to upper bounds for a one-step Kron model reduction as presented in Table~\ref{tab:table1}.
\begin{table}[!h]
    \centering
    \begin{tabular}{|c|c|c|c|}
    \hline 
    Removed node & $1$ & $2$ & $3$ \\ \hline 
      $2M_{ii}\sqrt{\pi_i^c\pi_i^o}$& $3.6978$  &  $0.3595$  &  $1.2017$ \\  
      $L_2$-norm error & $0.4075$ &   $0.0335$   & $0.1016$ \\
       \hline 
    \end{tabular}
    \caption{The computation result of upper bounds and $L_2$-norm error of Example~\ref{ex_2_1}.}
    \label{tab:table1}
\end{table}
From Table~\ref{tab:table1}, the upper bounds indeed guide us to choose which complex to be removed. Namely, small upper bound corresponds to small  $\mathcal{H}_\infty$-norm error. In this example, the smallest error is obtained by removing complex 2. The conservatism in the error bound is mainly due to the fact that a diagonal structure is enforced 
in the generalized Gramians in Definition~\ref{def:gen_gram}.
\end{example}

In practice, when we apply our Kron reduction method to a CRN, we need to truncate not only one complex. From the a priori upper bound as 
in Proposition~\ref{thm:upper_bound}, we can extend this bound for the truncation of a set of complexes. This upper bound is given in the following theorem. 
\begin{theorem}\label{thm:upper_bound_sum}
Consider an open SS CRN system $\Sigma$ as in \eqref{eq:CRN_i/o_SS} and the corresponding Kron reduced-order model 
$\hat{\Sigma}_r$ as in \eqref{eq:reduced_model_SS} with 
$r<n$ where a set of $n-r$ complexes that are not measured has been removed from the network through Kron reduction. 
Then for any input  $u \in \mathcal{L}_2[0,\infty)$ and initial condition $x(0)=0$ and $\hat{x}(0)=0$, the outputs $y$ and $\hat{y}_r$ satisfy
\begin{equation}\label{eq:bound_sum}
    \lVert y - \hat{y}_r \rVert_2 \leq 2 \left( \sum_{i=r+1}^{n} M_{ii} \sqrt{\pi_i^c \pi_i^o}\right) \lVert u \rVert_2,
\end{equation}
where $y$ and $\hat{y}_r$ are the outputs of the original and the reduced-order model, respectively,
$\pi_i^c$s and $\pi_i^c$s are the removed generalized controllability and observability Gramians, respectively, as in \eqref{eq:gen_Gramians}, and  
the scalar $M_{ii}$ is the $i$-th diagonal element of the matrix $M=-A^{-1}=(L+R)^{-1}$.
\end{theorem}
\begin{proof}
By triangular inequalities, we have that
\begingroup\footnotesize\begin{align}
    \lVert y - \hat{y}_r \rVert_2 &= \lVert y - \hat{y}_{n-1} + \hat{y}_{n-1} - \cdots + \hat{y}_{r+1}-\hat{y}_{r} \rVert_2 \nonumber \\
    & \leq \lVert y - \hat{y}_{n-1} \rVert_2 + \lVert \hat{y}_{n-1} - \hat{y}_{n-2} \rVert_2  + \cdots + \lVert \hat{y}_{r+1} - \hat{y}_{r} \rVert_2 \nonumber \\
    &=\sum_{i=r+1}^{n} \lVert \hat{y}_{i} - \hat{y}_{i-1} \rVert_2 \label{eq:sum_y}
\end{align}\endgroup
with $\hat{y}_n := y$. 
It is clear that each term in the summation \eqref{eq:sum_y} is an error of a one-step Kron reduction. From Proposition~\ref{prop:Gram_red}, it follows that the Gramians of system $\hat{\Sigma}_i$ (corresponding to the output $\hat{y}_i$) is given by $\mathrm{diag}(\pi^c_1,\ldots,\pi^c_i)$ and $\mathrm{diag}(\pi^o_1,\ldots,\pi^o_i)$. Moreover, since $M=-{A}^{-1}$ and $M$ admits partition $\eqref{eq:partition_M}$, $M_{ii}$ is the diagonal element corresponding to truncation of system $\Sigma_i$ to $\Sigma_{i-1}$ via Kron reduction. Therefore, from Proposition~\ref{thm:upper_bound},  each error $\hat{y}_i-\hat{y}_{i-1}$ satisfies the bound $\lVert \hat{y}_i - \hat{y}_{i-1} \rVert_2 \leq 2M_{ii} \sqrt{\pi_i^c \pi_i^o} \lVert u \rVert_2$. Hence we obtain \eqref{eq:bound_sum} as claimed. 
\end{proof}

\section{Numerical Examples 
}\label{sec:example}

In this section, we will evaluate numerically the efficacy of Kron reduction method and validate the results of previous sections in two open CRN. The first one corresponds to the mass-action Activated Sludge Model  (ASM) that describes the dynamics in Wastewater Treatment Plant and is based on the well-known ASM1 model from \cite{MH-00}. The second one is the McKeithan's T-cell receptor signal transduction model  \cite{TWM-95}.  

\subsection{Mass-action Activated Sludge Model 1}

\begin{figure}
\begin{center}
\begin{tikzpicture}[node distance={15mm}] 
\node (x1) {$x_1$}; 
\node (x2) [right of=x1] {$x_2$};
\node (x3) [right of=x2] {$x_3$}; 
\node (x4) [right of=x3] {$x_4$};
\node (x5) [right of=x4] {$S_{NH}$};
\node (x6) [below of=x3] {$X_P$};
\node (x7) [above of=x3] {$x_5$};
\draw[->] (x1) -- node[midway, above] {$k_1$} (x2); 
\draw[->] (x2) -- node[midway, above] {$k_2$} (x3);
\draw[->] (x4) -- node[midway, above] {$k_4$} (x3);
\draw[->] (x5) -- node[midway, above] {$D_{\text{in}}$} (x4);
\draw[->] (x3) to[out=130,in=50] node[midway, above] {$k_5$} (x1);
\draw[->] (x3) to[out=210,in=330] node[midway, below] {$k_3$} (x2);
\draw[->] (x2) -- node[midway, left] {$k_{\text{out},1}$} (x6); 
\draw[->] (x4) -- node[midway, right] {$k_{\text{out},2}$} (x6);
\draw[->] (x2) -- node[midway, right] {$k_6$} (x7); 
\draw[->] (x4) to[out=70,in=340] node[midway, right] {$k_{7}$} (x7);
\draw[->] (x7) -- node[midway, right] {$k_{8}$} (x4);
\end{tikzpicture}
\caption{The graph of open CRN of ASM1 used in the first example where the state variables $x_i$, $i=1,\ldots 5$ represent the readily biodegradable substrate, heterotrophs biomass, slowly biodegradable substrate, autotrophs biomass and particulate organic nitrogen, respectively. The inflow is given by the constant influx of ammonium.} 
\label{fig:7}
\end{center}
\end{figure}
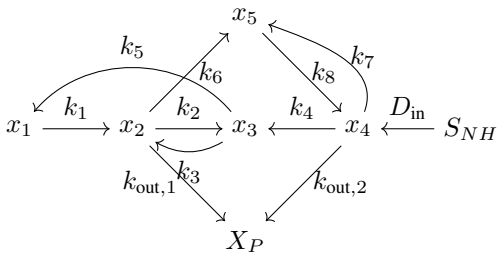

In the well-known Activated Sludge Model, which include ASM1, ASM2 and ASM3 \cite{MH-00}, the open CRN is dictated by general kinetics, as in \cite{BJ-12}. In this subsection, we  consider instead the ASM1 with mass-action kinetics, which are obtained by fixing the general kinetics part of ASM1 as constants, akin to the approach taken in \cite{AL-20}. Furthermore, we do not incorporate the nitrate and nitrite part of the ASM for simplifying the case and the corresponding open CRN is shown in Fig.~\ref{fig:7} where the inflow is given by the influx of ammonium $S_{NH}$ and the outflow is given by $k_{\text{out},1}x_2+k_{\text{out},2}x_4$ that represents particulate product $X_P$. The state variables $x_i$, $i=1,\ldots,5$ in Fig.~\ref{fig:7} are related to the variables in ASM1 as follows: $x_1=S_S$ (readily biodegradable substrate), $x_2=X_{BH}$ (heterotrophs biomass), $x_3=X_S$ (Slowly biodegradable substrate), $x_4=X_{BA}$ (autotrophs biomass) and $x_5=X_{ND}$ (particulate organic nitrogen). The time-series measurement of chemical oxygen demand (COD), which is commonly used to monitor activated sludge reactor, is given by the sum of both readily biodegradable substrate $x_1$ and slowly biodegradable substrate $x_3$. Correspondingly, the output matrix $C$ is given by $C=\bbm {1 & 0 & 1 & 0 &0}$.  

With a reference to the kinetics used in \cite{MH-00}, the rate constants $k_i$, $i=1,\ldots 8$, the constant inflow $D_{\text{in}}$ and outflow rate constant $k_{\text{out},1}, k_{\text{out},2}$, which appear typically in the Petersen matrix\footnote{In biochemistry literature, this term refers to a standard table of kinetics and the corresponding reaction rates in a given biochemical reaction system.} to describe the kinetics of biochemical processes, are given by  $k_1=k_3=\frac{Y_H}{1+b_H(1-Y_H(1-f_P))}$, $k_2=\frac{b_H}{1-f_P}$, $k_4=\frac{b_A}{1-f_P}$,  $k_5=\frac{1+b_H}{\mu_H-(1+b_H)}$, $k_6=\frac{b_A}{i_{xB}-f_Pi_{xP}}$, $k_7=\frac{b_H}{i_{xB}-f_Pi_{xP}}$, $k_8=\frac{Y_A}{(1+Y_Ai_{xB})(1+b_A)-Y_Ab_A(i_{xB}-f_Pi_{xP})}$, $k_{\text{out},1}=f_Pb_H$, $k_{\text{out},2}=f_Pb_A$ and $D_{\text{in}}=\frac{Y_A}{(1+Y_Ai_{xB})(1+b_A)-Y_Ab_A(i_{xB}-f_Pi_{xP})}$ where $Y_H$, $Y_A$, $f_P$, $b_H$, $b_A$, $\mu_H$, $i_{xP}$ and $i_{xB}$ are constant parameters as used in \cite{MH-00}. Using the numerical values in \cite{MH-00}, we have the following numerical values
\begin{equation*}
\begin{array}{ccccc}
k_1=0.54;&k_2=0.67;&k_3=0.54;&k_4=0.19;\\
k_5=0.37;&k_6=2.22;&k_7=7.64;&k_8=1.19;\\
k_{\text{out},1}=0.05;&k_{\text{out},2}=0.01;\text{and}&D_{\text{in}}=1.19.\\
\end{array}
\end{equation*}
Correspondingly, the matrices of the open SS CRN in \eqref{eq:CRN_i/o_SS} are given by 
\begin{align*}
    A&=\sbm{-0.54 & 0 & 0.37 & 0 & 0\\
            0.54 & -2.94 & 0.54 & 0 & 0\\
            0 & 0.67 & -0.91 & 0.19 & 0\\
            0 & 2.22 & 0 & -7.84 & 1.19\\
            0 & 0 & 0 & 7.64 & -1.19}, B=\sbm{0\\0\\0\\1.19\\0},\\
    C&=\sbm{1 & 0 & 1 & 0 &0}.
\end{align*}

\begin{figure}[t!]
\begin{center}
\input{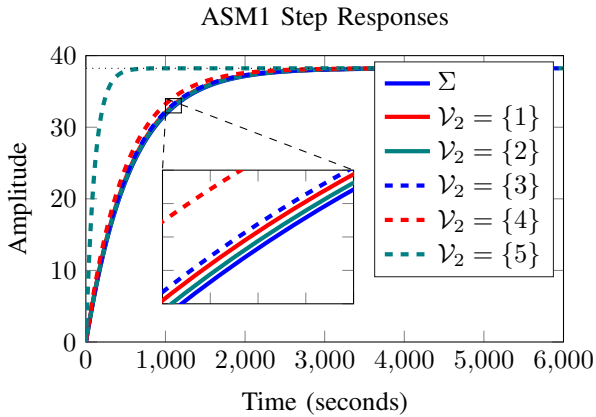}
\caption{The step response of the full-order open SS CRN ASM1 $\Sigma$ and the reduced-order ones via one-step Kron reduction step where the removed node is given by $\mathcal{V}_2$. }
\label{fig:8}
\end{center}
\end{figure}

Using the one-step model reduction as presented in the previous section, we can compute the upper bound of the output error as presented in 
Proposition~\ref{thm:upper_bound}. Whilst the ASM1 used in this example is not detailed-balanced, we can follow Remark \ref{re:detailed_balanced} so that the bound \eqref{eq:bound_y/u} in Proposition~\ref{thm:upper_bound} still holds seeing that \eqref{eq:sup_delta_w} is valid for the corresponding $M$. The computed upper bounds for the different removed node $x_i$, $i=1,\ldots,5$ are presented in Table \ref{tab:table2}. As given in this table, the removal of $x_5$  will give the largest output error. This observation is validated by the numerical simulation of step response of the original open CRN and of each of one-step Kron-reduced open CRN as shown in 
Figure~\ref{fig:8}. In this figure, $\Sigma$ refers to the step response of the original network, while the other plots refer to that of the reduced-order model where the node in $\mathcal V_2$ is removed through Kron reduction. In this figure, the convergence rate $\lambda_1$ of all reduced-order models is less than that of the full-order one, which is in accordance to the spectrum interlacing property as in Proposition \ref{prop:interlacing}. As shown also in the figure, the zero moment matching is achieved in all of these one-step Kron-reduced open CRNs following the result in Proposition \ref{prop1}.    
\begin{table}[!th]
    \centering
    \begin{tabular}{|c|c|c|c|c|c|}
    \hline 
    Removed complex & $x_1$ & $x_2$ & $x_3$ & $x_4$ & $x_5$ \\ \hline 
      $2M_{ii}\sqrt{\pi_i^c\pi_i^o}$& $1.3517$ & $0.0701$ & $0.2235$ & $0.9275$ & $6.0011$ \\
       \hline 
    \end{tabular}
    \caption{The upper bound computation of one-step Kron reduction to the mass-action kinetics ASM1 according to Proposition~\ref{thm:upper_bound}.}
    \label{tab:table2}
\end{table}





\subsection{Mass-action McKeithan's open CRN}

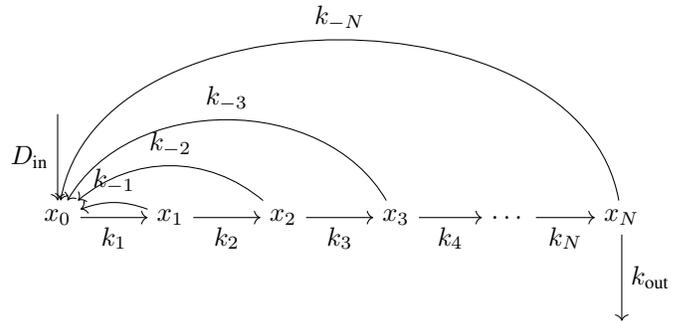
\begin{figure}
\begin{center}
\begin{tikzpicture}[node distance={15mm}] 
\node (x0) {$x_0$}; 
\node (x1) [right of=x0] {$x_1$}; 
\node (x2) [right of=x1] {$x_2$};
\node (x3) [right of=x2] {$x_3$}; 
\node (dot) [right of=x3] {$\dots$};
\node (xn) [right of=dot] {$x_N$};
\node (b) [above of=x0] {};
\node (a) [below of=xn] {};
\draw[->] (b) -- node[midway, left] {$D_{\text{in}}$} (x0);
\draw[->] (x0) -- node[midway, below] {$k_1$} (x1); 
\draw[->] (x1) to[out=160,in=20] node[midway, above] {$k_{-1}$} (x0); 
\draw[->] (x1) -- node[midway, below] {$k_2$} (x2);
\draw[->] (x2) to[out=140,in=40] node[midway, above] {$k_{-2}$} (x0); 
\draw[->] (x2) -- node[midway, below] {$k_3$} (x3);
\draw[->] (x3) to[out=120,in=60] node[midway, above] {$k_{-3}$} (x0);
\draw[->] (x3) -- node[midway, below] {$k_4$} (dot);
\draw[->] (dot) -- node[midway, below] {$k_N$} (xn);
\draw[->] (xn) to[out=100,in=80] node[midway, above] {$k_{-N}$} (x0);
\draw[->] (xn) -- node[midway, right] {$k_{\text{out}}$} (a);
\end{tikzpicture}
\caption{McKeithan's network with inflow and outflow.} 
\label{fig:4}
\end{center}
\end{figure}

In this subsection, we evaluate the applicability and efficacy of our main results to the McKeithan's open CRN (see \cite{TWM-95}). This model was developed to describe the selectivity of T-cell interactions. With reference to Fig.~\ref{fig:4}, $x_0$ represents a T-Cell receptor and peptide-major histocompatibility complex. For every $i=1,2,...,N$ $x_i$ represents various intermediate complex in the phosphorylation and other intermediate modification of the T-cell receptor; $k_i$ represents the rate constant of the $i^{\text{th}}$ step of the phosphorylation and $k_{-i}$ is the dissociation rate of the $i^{\text{th}}$ complex.

For this example, we use the following numerical values (as used in \cite{SR-13}) with $N=20$ 
\begin{equation*}
\begin{array}{ccccc}
k_1=52;&k_2=49;&k_3=41;&k_4=39;\\
k_5=37;&k_6=34;&k_7=31;&k_8=29;\\
k_9=25;&k_{10}=19;&k_{11}=16;&k_{12}=21;\\
k_{13}=20;&k_{14}=19;&k_{15}=18;&k_{16}=15;\\
k_{17}=24;&k_{18}=13;&k_{19}=7;&k_{20}=5;\\
k_{-1}=13;&k_{-2}=29;&k_{-3}=0.16;&k_{-4}=1.4;\\
k_{-5}=2.3;&k_{-6}=2;&k_{-7}=0.19;&k_{-8}=0.33;\\
k_{-9}=0.94;&k_{-10}=0.67;&k_{-11}=0.31;&k_{-12}=0.21;\\
k_{-13}=3;&k_{-14}=5;&k_{-15}=1;&k_{-16}=11;\\
k_{-17}=0.8;&k_{-18}=7;&k_{-19}=1;&k_{-20}=17;\\
\end{array}
\end{equation*}
and we introduce an inflow $D_{\text{in}}=1$ and an outflow $k_{\text{out}}=10$. In accordance with Fig.~\ref{fig:4}, the input and output matrices $B \in \mathbb{R}^{21 \times 1}$ and $C \in \mathbb{R}^{1 \times 21}$  are given by
\begin{equation}\nonumber 
    B=\bbm{1 & 0 & \cdots & 0}^T \ \text{and } \ C=\bbm{0 & \cdots & 0 & 1},
\end{equation}
respectively.

Let us  consider the application of one-step Kron-reduction to this network. Although the McKeithan network is not detailed-balanced, following Remark \ref{re:detailed_balanced}, the bound \eqref{eq:bound_y/u}  in Proposition~\ref{thm:upper_bound} still holds since  \eqref{eq:sup_delta_w} is valid for the corresponding $M$ in this example. The computed upper bound \eqref{eq:bound_y/u} 
and the computed $\mathcal{H}_\infty$-norm of the model discrepancies are given in Table~\ref{tab:result}. Table~\ref{tab:result} shows that the upper bounds computed by \eqref{eq:bound_y/u} enable us to select which node to be removed to obtain smaller error. Namely, the removal of nodes with small error bound corresponds to small approximation error. However, as in Example~\ref{ex:gramian_based}, the upper bounds are conservative because the Gramians are enforced to be diagonal. 

\begin{table}[!ht]
    \centering
    \begin{tabular}{|c|c|c|}
    \hline 
    Removed node & Bound \eqref{eq:bound_y/u} & $\frac{\lVert y -\hat{y} \rVert_2}{ \lVert u \rVert_2}$ ($\times 10^{-3}$) \\
    \hline 
   21  &  0.2436  &  0.4283 \\
   17  &  0.5249  &  1.1678\\
    3  &  0.9024  &  2.1484\\
   19  &  0.9229  &  1.9059\\
    4  &  0.9486  &  2.2514\\
    5  &  0.9636  &  2.2874\\
   18  &  0.9811  &  2.0534\\
    6  &  0.9830  &  2.3325\\
    7  &  1.0148  &  2.4048\\
   15  &  1.0570  &  2.4492\\
    8  &  1.0803  &  2.5576\\
   16  &  1.2222  &  2.7646\\
    9  &  1.2377  &  2.9387\\
   14  &  1.2641  &  2.9795\\
    2  &  1.2892  &  3.0969\\
   12  &  1.3359  &  3.1595\\
   20  &  1.3492  &  2.2954\\
   13  &  1.3896  &  3.2869\\
    1  &  1.5374  &  3.7140\\
   10  &  1.5533  &  3.7112\\
   11  &  1.7753  &  4.2513\\
   \hline 
    \end{tabular}
    \caption{Computation result of (increasingly sorted) error bounds \eqref{eq:bound_y/u} for
    one step Kron-reductions. These bounds are compared with the actual $\mathcal{H}_\infty$-norm errors. }
    \label{tab:result}
\end{table}

\begin{figure}[!t]
    \centering
    \input{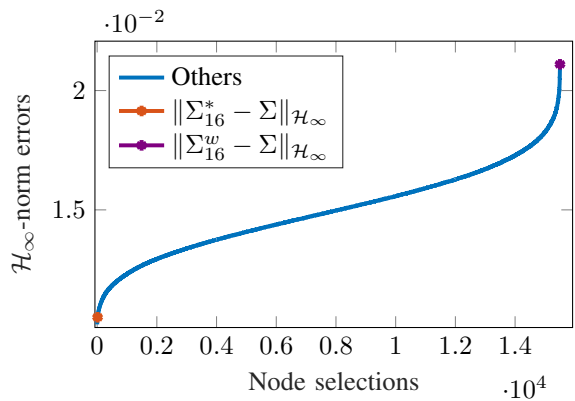}
    \caption{Comparison of model reduction error using this Gramian-based node selection compared with all other possible node selections. 
    The system $\hat{\Sigma}_{16}^*$ denotes the reduced-order model of order $16$ by truncating nodes $17,3,19,4,5$. The $\mathcal{H}_\infty$-norm error of this Gramian-based nodes selection is $0.0105$ marked by the red asterisk symbol, where the smallest error is $0.0102$.
     In addition, $\hat{\Sigma}_{16}^{\mathrm{w}}$ denotes the reduced order model with the `worst' node selection, i.e., largest error.}
    \label{fig:comparison_to_all}
\end{figure}

For showing the applicability of these bounds, we first aim at obtaining 
reduced-order model of order $r=16$, i.e., $5$ nodes are removed from the network. According to Table~\ref{tab:result}, the truncation of nodes
$17,3,19,4,5$ will lead to 
 small approximation error. 
The resulting approximation error of reduced-order model by removing  these nodes is 
\begin{equation}\label{eq:appr_errorMckeithan}
    \frac{\lVert y - \hat{y} \rVert_2}{\lVert u \rVert_2} = 0.0105.
\end{equation}
Here, we compare the $\mathcal{H}_\infty$ norm error of this Gramian-based selection to all other possible nodes removal. Note that choosing $5$ from $20$ nodes gives us $15,504$ possible five-node combinations. We present the comparison of our result with respect to all other possible five-node combinations in Figure~\ref{fig:comparison_to_all}.
Among all combinations, the result from Gramian-based combination as in \eqref{eq:appr_errorMckeithan} is not the smallest but still gives a result that is very close to the optimal one.

\begin{figure}[!t]
    \centering
    \input{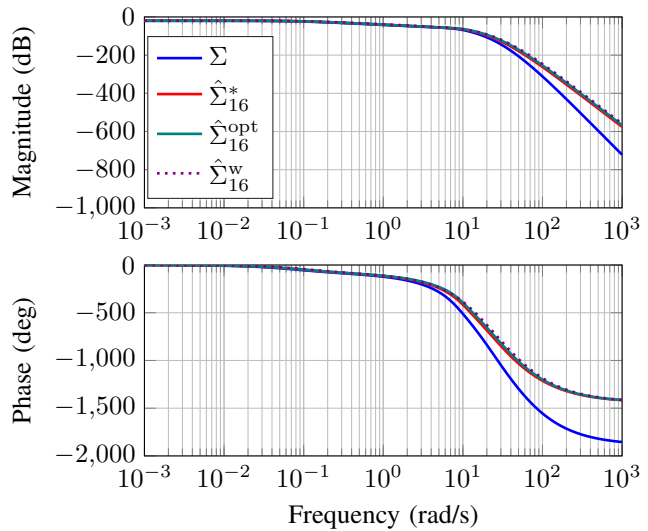}
    \caption{Bode diagrams of 
    the original system denoted by $\Sigma$,
    the reduced order model by truncating nodes $17,3,19,4,5$ denoted by $\hat{\Sigma}_{16}^*$, the reduced-order model of order $16$ (removing $5$ nodes) with minimum error denoted by $\hat{\Sigma}_{16}^{\mathrm{opt}}$ and the reduced-order model with the largest error $\hat{\Sigma}_{16}^{\mathrm{w}}$.}
    \label{fig:bode16}
\end{figure}

\begin{figure}[!t]
   \centering
    \input{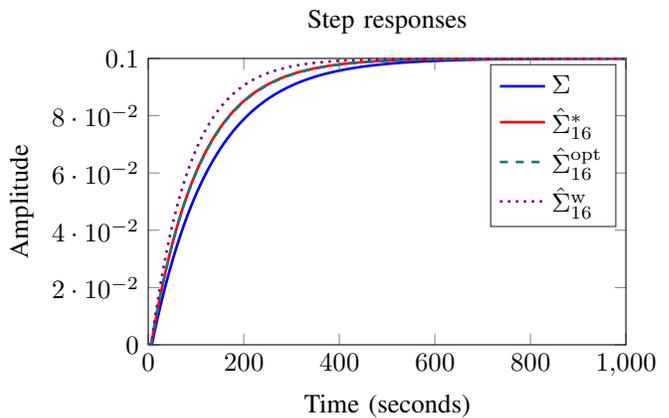}
    \caption{Comparison of the step responses of the original system denoted by $\Sigma$ and the reduced-order models of McKeithan network. }
    \label{fig:step}
\end{figure}

In order to illustrate the resulting reduced-order model $\hat{\Sigma}^*_{16}$, we also present both frequency-domain by means of a Bode diagrams and time-domain responses in Figures~\ref{fig:bode16} and \ref{fig:step}, respectively. Both the Bode diagrams and step responses show that the Gramian-Kron-based reduced-order model almost coincides to the Kron-based reduced-order model with the optimal nodes combination. Similar observation to the previous example, the convergence rate of all reduced-order models in Figure \ref{fig:step} is less than that of the full-order one shown in solid-blue line, which follows the spectrum interlacing property in Proposition \ref{prop:interlacing}. Another noteworthy observation is that even the reduced-order model of order $16$ with the largest error (denoted by $\hat{\Sigma}_{16}^{\mathrm{w}}$ in Figure~\ref{fig:comparison_to_all}) still results in a good approximation.

Aside from the comparison of Kron-based reduced-order models, all reduced-order models in this example affirm the zero-moment matching property as presented in Proposition~\ref{prop1}.  Figure~\ref{fig:bode16} shows that all reduced-order models match the moment of the original system at frequency $0$ rad/s. The time-domain response is shown in Figure~\ref{fig:step} where the steady state responses of the original model and all reduced-order models coincide with each other.

\begin{figure}[t!]
    \centering
    \input{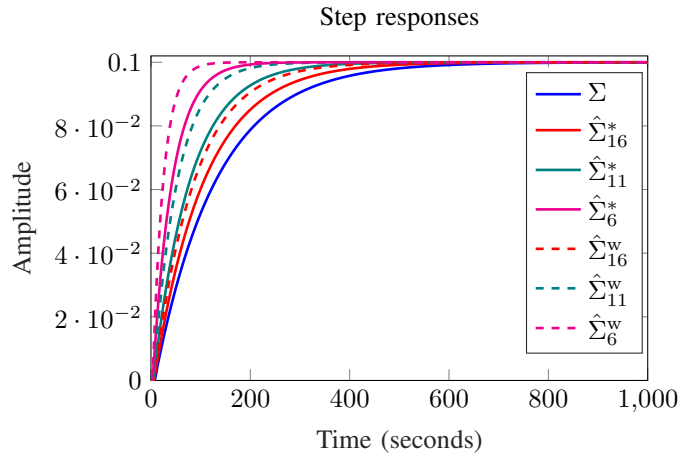}
    \caption{Step responses of reduced-order models of McKeithan network with varying number of of truncated nodes.}
    \label{fig:step_var}
\end{figure}

In the following, we will evaluate the efficacy of Kron-based model reduction with larger truncation set that leads to models with lower order than before. We consider the removal of $10$ nodes and of $15$ nodes following the three different cases as before. For the removal of $10$ nodes, we firstly compute the reduced-order model by removing $10$ nodes according to the Gramian upper bound in Table~\ref{tab:result}. Secondly, we determine the optimal $10$ nodes that give the best $\mathcal{H}_\infty$-norm error. Thirdly, we find the $10$ nodes that give the worst  $\mathcal{H}_\infty$-norm error. 
The first one corresponds to the removal of nodes $3, 4, 5, 6, 7, 8, 15, 17, 18, 19$, the second one corresponds to the removal of $3, 4, 5, 6, 7, 15, 17, 18 ,19 ,20$ and, finally, the third one corresponds to removal of $1, 2, 8, 9, 10, 11, 12, 13, 14, 16$. Similarly, for the removal of $15$ nodes that leads to a reduced-order model of order $6$, we consider the same three different cases as above. From Table~\ref{tab:result}, we \emph{keep} the nodes set $1, 10,    11, 13, 20, 21$. For the best and the worst nodes combination, we keep the nodes set $1, 10, 11, 12, 13,    21$ and $3, 17, 18, 19, 20, 21$, respectively. The comparison of these reduced-order models is presented in  Table~\ref{tab:comparison_all}. As an illustration, we also present the step responses of the Gramian-based and the worst truncation reduced-order models with varying number of truncated nodes in Figure~\ref{fig:step_var}. In this figure, the step response of Gramian-Kron based reduced-order models approximate well the full-order one, and the convergence rate of all reduced-order models follows the spectrum interlacing properties in Proposition \ref{prop:interlacing}. 


\begin{table}[!ht]
    \centering
    \begin{tabular}{c|c|c|c}
         &  $\lVert\hat{\Sigma}_{16}-\Sigma \rVert_\infty $ & $\lVert \hat{\Sigma}_{11}-\Sigma\rVert_\infty$ & $\lVert \hat{\Sigma}_{6}-\Sigma \rVert_\infty$ \\
         \hline 
       Optimal & $0.0102$ & $0.0258$         & $0.0516$ \\
       reduced-order model  & & & \\
     \hline 
     Gramian-based   & $0.0105$ & $0.0264$ &$0.0540$\\ reduced-order model & & & \\
     \hline 
     The `worst'  & $0.0221$ & $0.0452$ & $0.0731$ \\
     reduced-order model & & & \\
     \hline
    \end{tabular}
    \caption{Comparison of $\mathcal{H}_\infty$-norm error of reduced order model via Kron reduction with variation of nodes selection and number of truncated nodes.}
    \label{tab:comparison_all}
\end{table}

\section{Conclusions}\label{sec:conclusion}

In this paper, we have presented  Kron reduction approach to get  reduced-order models of open chemical reaction networks (CRN) with mass-action kinetics. We show a number of systems properties that are inherited by the reduced-order model, namely, the open CRN structure, the zero-moment matching property, the spectrum interlacing property, and the upper bound of the approximation error via generalized Gramian approach. The latter property has allowed us to guide systematically the selection of removed nodes/species via Kron reduction. The applicability and efficacy of our method and analysis have been shown in two well-known biochemical reaction networks: the activated sludge model 1 and the McKeithan's T-cell receptor model.   

\section*{Acknowledgement}
We would like to thank Bart Besselink for discussions on the topic of Section~\ref{sec:Gramians} of this paper.

\bibliographystyle{IEEEtran}
\bibliography{Reference.bib}
\end{document}